\documentclass[]{aastex6}

\usepackage{amsmath}
\usepackage{setspace}
\slugcomment{To appear in Planetary and Space Science (doi:10.1016/j.pss.2019.104825)}
\begin{document}

%% LaTeX will automatically break titles if they run longer than
%% one line. However, you may use \\ to force a line break if
%% you desire.

\title{The Morphological, Elastic, and Electric Properties of Dust Aggregates in Comets: A Close Look at COSIMA/Rosetta's Data on Dust in Comet 67P/Churyumov-Gerasimenko}

%% Use \author, \affil, plus the \and command to format author and affiliation 
%% information.  If done correctly the peer review system will be able to
%% automatically put the author and affiliation information from the manuscript
%% and save the corresponding author the trouble of entering it by hand.
%%
%% The \affil should be used to document primary affiliations and the
%% \altaffil should be used for secondary affiliations, titles, or email.

%% Authors with the same affiliation can be grouped in a single
%% \author and \affil call.
\author{Hiroshi Kimura\altaffilmark{1}}
\affil{Planetary Exploration Research Center (PERC), Chiba Institute of Technology, Tsudanuma 2-17-1, Narashino, Chiba 275-0016, Japan}
\affil{Max Planck Institute for Solar System Research, Justus-von-Liebig-Weg 3, 37077 G\"{o}̈ttingen, Germany}

\author{Martin Hilchenbach}
\author{Sihane Merouane}
\author{John Paquette}
\author{Oliver Stenzel}
\affil{Max Planck Institute for Solar System Research, Justus-von-Liebig-Weg 3, 37077 G\"{o}̈ttingen, Germany}

%\author{COSIMA team}

%% Use the \and command so offset the last author.
%\and

%\author{Jeff Lewandowski\altaffilmark{5}}
%\affil{IOP Publishing, Washington, DC 20005}

%% Notice that each of these authors has alternate affiliations, which
%% are identified by the \altaffilmark after each name.  Specify alternate
%% affiliation information with \altaffiltext, with one command per each
%% affiliation.

\altaffiltext{1}{hiroshi\_kimura@perc.it-chiba.ac.jp}

%% Mark off the abstract in the ``abstract'' environment. 
\begin{abstract}
%\linenumbers

The Cometary Secondary Ion Mass Analyzer (COSIMA) onboard ESA's Rosetta orbiter has revealed that dust particles in the coma of comet 67P/Churyumov-Gerasimenko are aggregates of small grains.
We study the morphological, elastic, and electric properties of dust aggregates in the coma of comet 67P/Churyumov-Gerasimenko using optical microscopic images taken by the COSIMA instrument.
Dust aggregates in COSIMA images are well represented as fractals in harmony with morphological data from MIDAS (Micro-Imaging Dust Analysis System) and GIADA (Grain Impact Analyzer and Dust Accumulator) onboard Rosetta.
COSIMA's images, together with the data from the other Rosetta's instruments such as MIDAS and GIADA do not contradict the so-called rainout growth of $10~\micron$-sized particles in the solar nebula.
The elastic and electric properties of dust aggregates measured by COSIMA suggest that the surface chemistry of cometary dust is well represented as carbonaceous matter rather than silicates or ices, consistent with the mass spectra, and that organic matter is to some extent carbonized by solar radiation, as inferred from optical and infrared observations of various comets.
Electrostatic lofting of cometary dust by intense electric fields at the terminator of its parent comet is unlikely, unless the surface chemistry of the dust changes from a dielectric to a conductor.
Our findings are not in conflict with our current understanding of comet formation and evolution, which begin with the accumulation of condensates in the solar nebula and follow with the formation of a dust mantle in the inner solar system.

\end{abstract}

%% Keywords should appear after the \end{abstract} command. 
%% See the online documentation for the full list of available subject
%% keywords and the rules for their use.

%% From the front matter, we move on to the body of the paper.
%% Sections are demarcated by \section and \subsection, respectively.
%% Observe the use of the LaTeX \label
%% command after the \subsection to give a symbolic KEY to the
%% subsection for cross-referencing in a \ref command.
%% You can use LaTeX's \ref and \label commands to keep track of
%% cross-references to sections, equations, tables, and figures.
%% That way, if you change the order of any elements, LaTeX will
%% automatically renumber them.

%% We recommend that authors also use the natbib \citep
%% and \citet commands to identify citations.  The citations are
%% tied to the reference list via symbolic KEYs. The KEY corresponds
%% to the KEY in the \bibitem in the reference list below. 

\section{Introduction} \label{sec:intro}

The formation and evolution of comets are a long-standing issue for planetary scientists, while a study on the physical and chemical properties of cometary dust provides an important clue to correct understanding of comets.
In the 20th century, the most popular scenario for the formation of comets was the accumulation of presolar interstellar grains that were preserved in the solar nebula \citep[e.g.,][]{cameron1975,greenberg1998}.
As a consequence, an expected picture of cometary dust is an aggregate particle of submicrometer-sized elongated interstellar grains consisting of a silicate core, an organic inner mantle, and an ice outer mantle \citep{greenberg-hage1990}.
Thanks to laboratory analyses of interplanetary dust particles (IDPs) collected in the stratosphere of the Earth, our knowledge of cometary dust has advanced over the decades \citep{brownlee1985}.
There is consensus among experts on IDPs that a chondritic porous (CP) subset of IDPs is of cometary origin and bears compositionally strong resemblance to dust in comet 1P/Halley measured in situ \citep[e.g.,][]{jessberger1999}.
The major constituent of CP IDPs is amorphous silicate called GEMS (glass embedded with metal and sulfides) that is isotopically homogeneous, compositionally non-solar, and submicrometer-sized \citep{bradley1994}.
GEMS grains were claimed to be of interstellar origin, because they share common properties with interstellar grains such as their submicrometer-size, amorphous structure, and infrared spectra \citep{bradley-et-al1999}.
However, it turned out from thorough laboratory examination of GEMS early this century that GEMS grains have complementary compositions to crystalline grains and thus both formed from the same reservoir of the solar nebula \citep{keller-messenger2011,keller-messenger2013}\footnote{There are presolar GEMS grains with isotopic anomalies in CP IDPs, but they occupy only a few percent of total number of GEMS grains present in CP IDPs \citep{keller-messenger2011}.}.
Accordingly, we are now aware that comets are made out of condensates in the solar nebula rather than presolar interstellar grains, contrary to the most popular scenario of comet formation in the 20th century.
Therefore, thorough investigation of cometary dust will help us to understand the formation of pristine materials in the solar system and the early stages of planet formation.

CP IDPs are aggregate particles consisting of submicrometer-sized Mg-rich crystalline silicates and GEMS glued together by carbonaceous material \citep{keller-et-al2000,flynn-et-al2013}.
Such an aggregate structure is a natural consequence of dust growth by coagulation in the solar nebula, according to a model for the formation of comets \citep{weidenschilling1997}.
It should be noted that the relatively low degree of fluffiness of CP IDPs is not necessarily a typical example of cometary dust, because only compact endmembers of cometary dust are transported to the Earth by the Poynting-Robertson effect \citep{kimura-et-al2016}.
Owing to a stronger radiation pressure force on a fluffy particle than a compact particle of the same mass, the former tends to have a difficulty of staying in a bound orbit around the sun \citep[cf.][]{kresak1976}.
The Deep Impact mission to comet 9P/Tempel 1 revealed that the radiation pressure on dust particles excavated by a projectile into the comet is strong enough to place the particles in a hyperbolic orbit \citep[see][]{kobayashi-et-al2013}.
It is interesting to note that a dust mantle of a short-period comet is most likely composed of relatively compact aggregates with a radius of tens of microns \citep{li-greenberg1998,kolokolova-et-al2007,yamamoto-et-al2008,kobayashi-et-al2013}.
Therefore, we may expect that in-situ measurements of cometary dust give new insights into not only the composition, but also  the morphology of aggregate particles in comets.

A large number of dust particles from comet 67P/Churyumov-Gerasimenko have been collected by the Cometary Secondary Ion Mass Analyzer (COSIMA) onboard the ESA's Rosetta orbiter \citep{langevin-et-al2016,merouane-et-al2017}.
COSIMA is equipped with an optical microscope called COSISCOPE that acts as a powerful tool to investigate the morphology of dust particles and their collisional and electrostatic characteristics.
We shall explore the data obtained by Rosetta/COSIMA and discuss them together with the other Rosetta's data from a theoretical point of view in terms of physical and chemical properties of cometary dust.

\section{Theoretical Backgrounds} %\label{sec:image}

\subsection{Coagulation of Dust Aggregates in the Solar Nebula} %\label{sec:image}

In the early stage of comet formation, dust aggregates grow under ballistic cluster-cluster aggregation (BCCA) process, because the motion of dust particles is controlled by Brownian motion where dust particles of similar size hit and stick each other \citep{weidenschilling-et-al1989}.
Once aggregate particles grow to the size of 10--$10^2~\micron$, they begin to settle toward the central plane of the solar nebula (``rainout'') and inevitably sweep up smaller aggregate particles \citep{weidenschilling1997}.
Therefore, we may expect that the rainout particles are well characterized by aggregate particles grown under ballistic particle-cluster aggregation (BPCA) process.

A plausible scenario for the formation of planetesimals in the solar nebula requires that dust aggregates grew against mutual collisions even at a relative velocity of $50~\mathrm{m~s^{-1}}$ at 30 au from the Sun \citep{weidenschilling-cuzzi1993,weidenschilling1997}.
It is not impossible for aggregates of silicate grains or aggregates of ice grains to stick each other at a collision velocity of $50~\mathrm{m~s^{-1}}$, but it is much easier for aggregates of organic grains to grow in the solar nebula \citep{kimura-et-al2015}.
As a result, it seems plausible that the surface of submicrometer-sized grains originally consists of organic matter when the grains grew to comets by coagulation in the solar nebula.

\subsection{The Concept of Fractal Geometry} 

It has been well known that the structure of aggregate particles grown under the BCCA and BPCA processes is well characterized by the concept of fractals \citep{meakin-donn1988}.
The number $N$ of constituent grains with radius $r_0$ and the characteristic radius $r_\mathrm{c}$ of a fractal aggregate define a fractal dimension\footnote{This is also called the Hausdorff dimension.} $D$ of the aggregate:
\begin{linenomath*}
\begin{eqnarray}
N = k_0 {\left({\frac{r_\mathrm{c}}{r_0}}\right)}^{D}, 
\label{fractal}
\end{eqnarray} 
\end{linenomath*}
where $k_0$ is a proportionality constant of order unity.
Hereafter we assume $\log k_0 = -0.576 D + 0.915$ \citep{kimura-et-al1997}.
BCCA and BPCA particles are known to have their fractal dimensions $D \approx 2$ and $D \approx 3$, respectively, if their constituent grains are assumed to hit and stick on contact.
In reality, the fractal dimension $D$ of BCCA particles depends on the the relative velocity of mutual collisions where low velocities with no restructuring indicate $D \approx 2$ and high velocities with the maximum compression result in $D \approx 2.5$ \citep{wada-et-al2008}.
We adopt the following definition for the porosity $po$ of an aggregate particle proposed by \citet{mukai-et-al1992}:
\begin{linenomath*}
\begin{eqnarray}
po = 1 - k_0 {\left({\frac{r_\mathrm{c}}{r_0}}\right)}^{D-3} ,
\end{eqnarray} 
\end{linenomath*}
with 
\begin{linenomath*}
\begin{eqnarray}
r_\mathrm{c} = \sqrt{\frac{5}{3}} r_\mathrm{g} ,
\end{eqnarray} 
\end{linenomath*}
where $r_\mathrm{g}$ denotes the radius of gyration.
The porosity of aggregate particles decreases with their size if $D < 3$, while the porosity of BPCA particles with $D \approx 3$ is approximately constant at $po \approx 0.85$, if their constituent grains hit and stick on contact \citep{mukai-et-al1992,kimura-et-al2016}.
This is in accordance with $po = 0.87$ derived for the outermost dust mantle of comet 67P/Churyumov-Gerasimenko from the so-called Hapke's modeling of a spectral variation in the reflectance of the nucleus \citep{fornasier-et-al2015}.

\subsection{Fractal Dimensions} 

The fractal dimension $d$ in a two-dimensional (2-D) Euclidean plane is related to the fractal dimension $D$ in three-dimensional (3-D) Euclidean space as follows \citep{meakin1991}:
\begin{linenomath*}
\begin{eqnarray}
  d = 
  \begin{cases}
    2 & \mathrm{if}~D \ge 2 , \\
    D & \mathrm{if}~D < 2 .
  \end{cases}
\label{fractal-dimension-in-2D}
\end{eqnarray} 
\end{linenomath*}
Therefore, for BCCA particles with $D \approx 2$ and BPCA particles with $D \approx 3$ in 3-D Euclidean space, their projected images onto a two-dimensional Euclidean plane are both characterized by a fractal dimension $d \approx 2$.
It is not straightforward to derive the fractal dimension $D$ in three-dimensional Euclidean space from two-dimensional projections of fractal aggregates \citep{maggi-winterwerp2004}.

We are aware that the fractal dimension of an aggregate particle slightly depends on the way how the fractal dimension is measured, although different dimensions could be generalized.
It should be noted that the correlation dimension $d_2$ might be smaller than the box-counting dimension $d_0$, which is 
defined by
\begin{linenomath*}
\begin{eqnarray}
d_0 = \frac{\log{n({\varepsilon})}}{\log{({1/\varepsilon}})} ,
\end{eqnarray} 
\end{linenomath*}
where $\varepsilon$ is the box size and $n({\varepsilon})$ is the minimum number of boxes that is required to cover the projected area of the aggregate particle.
The box-counting dimension is also called the capacity dimension and usually equivalent to the Hausdorff dimension, namely, $d_0 = d$ in 2-D Euclidean space.
Accordingly, aggregate particles with $d_2 < 2$ in a 2-D Euclidean plane are not necessarily associated with a low fractal dimension of $D < 2$ in 3-D Euclidean space.

\subsection{Processing of Dust Aggregates in a Dust Mantle} 

When comets evolve under solar radiation, the surface of the comets is processed and form the so-called dust mantle, a layer of dust particles devoid of volatile ices \citep{prialnik-et-al2004,yamamoto-et-al2008}.
Optical and infrared photo-polarimetric observations of dust in comets revealed that dust paricles in short-period comets are physically and chemically processed during the formation of a dust mantle \citep{kolokolova-et-al2007}.
Infrared spectra of cometary dust suggest that the organic refractory component of the dust is not intact, but is to some extent carbonized, resulting in the formation of amorphous carbon \citep{li-greenberg1998,kimura2014}\footnote{Here the term ``carbonization'' is used to indicate the loss of H, N, and O from the organic refractory component of comet dust by photolysis, radiolysis, and chemical reactions \citep[see][]{roessler-nebeling1987,jenniskens-et-al1993}}.
Therefore, it is reasonable to assume that the surface of submicrometer-sized grains consists of amorphous carbon to a large extent rather than pristine organic matter.

During the formation of a dust mantle on the surface of comets, dust particles with a high porosity, namely, a low fractal dimension are selectively ejected from the surface by gas drag, owing to their high cross-section to mass ratios.
Moreover, dust particles with a low porosity, namely, a high fractal dimension tend to fall back on the surface of a comet and thus most likely elevate the fractal dimension of dust particles in the surface of the comet as a result of inelastic collisions.
Therefore, it is reasonable to assume that dust particles in a dust mantle of a comet are composed of aggregate particles with relatively compact structures, in other words, high fractal dimensions, compared with those in the inner nucleus of the comet \citep[see][]{yamamoto-et-al2008}.

\section{Interpretation of COSIMA's data}

We shall explore what COSIMA's data as well as the other Rosetta's results tell us the morphological, elastic, and electric properties of cometary dust on the assumption that the morphology of the dust is well represented as a fractal.

\subsection{Morphologies}

\begin{figure}
\epsscale{1.0}
\plottwo{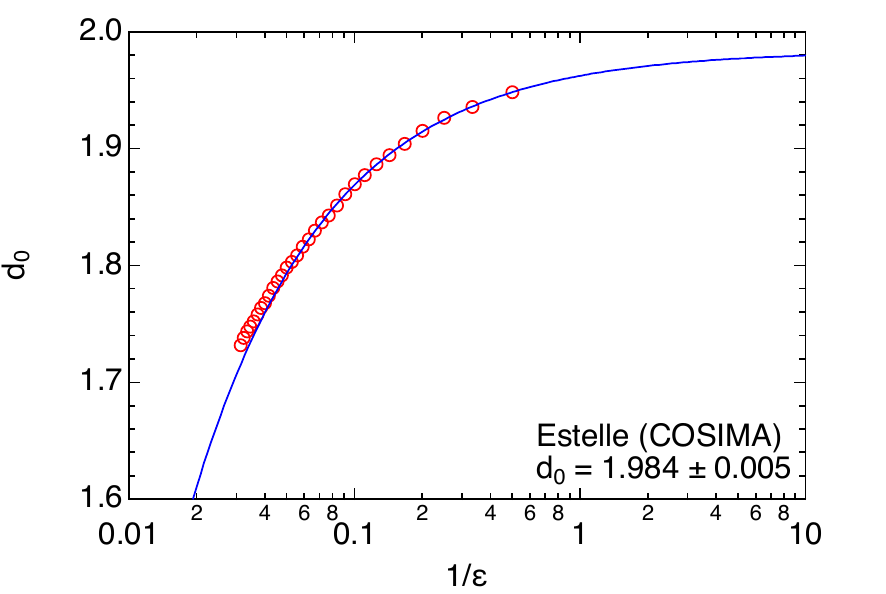}{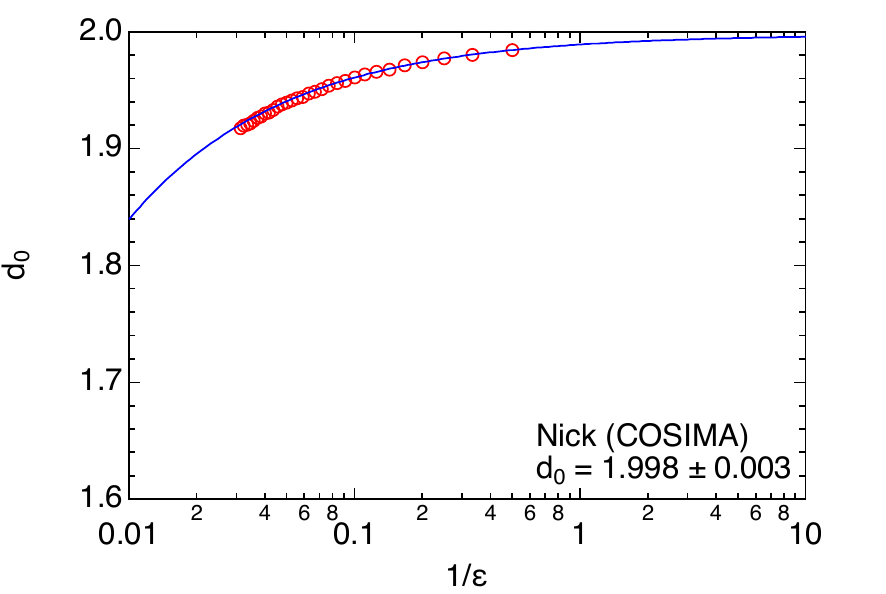}
\caption{The box-counting dimension $d_0(\varepsilon)$ (red circles) of the $100~\micron$-sized aggregates Estelle (left) and Nick (right) versus the inverse of the maximum box size $\varepsilon$.
The fitting curve (blue line) gives the box-counting dimension of Estelle and that of Nick, namely, the asymptotic values of $\lim_{\varepsilon \to 0} d_0(\varepsilon) = 1.984 \pm 0.005$ for Estelle and $\lim_{\varepsilon \to 0} d_0(\varepsilon) = 1.998 \pm 0.003$ for Nick.  \label{fig1}}
\end{figure}
COSISCOPE images of dust particles collected on the target of COSIMA revealed that the particles are an assemblage of subunits with an apparent radius of $\sim 10~\micron$ \citep{hornung-et-al2016,hilchenbach-et-al2017}.
Although COSISCOPE was not able to image submicrometer-sized grains expected as the constituent of dust aggregates in comets because of its resolution of $14~\micron$, the subunits do not seem to have broken up into minute fragments.
Therefore, COSIMA's data may indicate that the so-called rainout began at aggregate size of $\sim 10~\micron$ expected from a model for the formation of comets \citep[see][]{weidenschilling1997}.
By applying a box-counting algorithm to the COSISCOPE image of a $100~\micron$-sized aggregates named Estelle and Nick shown in \citet{langevin-et-al2016}, we obtain a box-counting dimension $d_0 = 1.984 \pm 0.005$ for the former and $d_0 = 1.998 \pm 0.003$ for the latter (see Fig.~\ref{fig1})\footnote{{We use the Java-based image processing program {\sf ImageJ} to obtain a box-counting dimension with the smallest box size being one pixel of images, namely, $14~\micron$.} When we perform curve fittings to box-counting dimensions, we restrict the data points to small box sizes of $1/\varepsilon \ge 0.1$.}.
The fractal dimension $d_0 \simeq 2$ in 2-D Euclidean space indicates that at least Estelle and Nick are fractal aggregates whose fractal dimensions in 3-D Euclidean space lie in the range of $D = 2$--$3$ (see Eq.~(\ref{fractal-dimension-in-2D})).
Therefore, COSISCOPE images are still in line with the rainout process in which aggregate particles are characterized by $D \approx 3$, although $d_0 \approx 2$ does not necessarily mean $D \approx 3$.

\begin{figure}
\epsscale{0.5}
\plotone{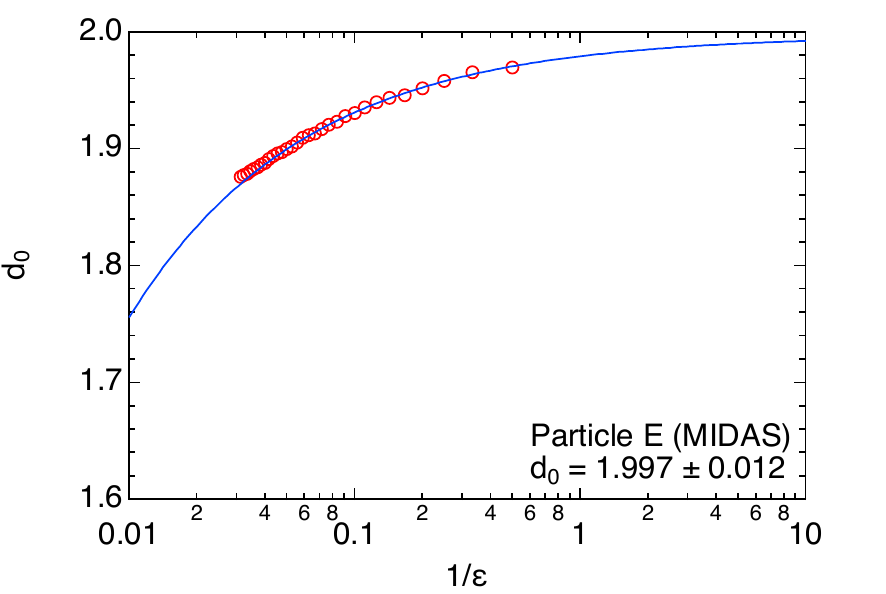}
\caption{The box-counting dimension $d_0(\varepsilon)$ (red circles) of the $100~\micron$-sized aggregate particle E (MIDAS) versus the inverse of the maximum box size $\varepsilon$.
The fitting curve (blue line) gives the box-counting dimension of particle E, namely, the asymptotic value of $\lim_{\varepsilon \to 0} d_0(\varepsilon) = 1.997 \pm 0.012$.  \label{fig2}}
\end{figure}
\citet{mannel-et-al2016} derived the correlation dimension of $d_2 \approx 1.70 \pm 0.10$ for a $10~\micron$-sized particle E from its 2-D projected image using a topographic image of the particle taken by MIDAS (Micro-Imaging Dust Analysis System) onboard Rosetta.
Since a correlation dimension $d_2$ could be smaller than a Hausdorff dimension $d$, it is not clear whether the particle E has a fractal dimension $D < 2$ in three-dimensional Euclidean space.
They also determined a Hausdorff dimension of the particle E to be $d \approx 1.76 \pm 0.29$ in a two-dimensional Euclidean plane, but the large uncertainty prevents justification of the claim that the particle E has a smaller fractal dimension than $d \approx 2.0$.
Moreover, the number of grains in the particle E is only 112 in two-dimensional Euclidean space and the fractal dimension might be underestimated due to the small number of grains used in their analysis.
We derive a box-counting dimension of $d_0 = 2.00 \pm 0.01$ from the 2-D projected image of the particle E in \citet{bentley-et-al2016}, as shown in Fig.~\ref{fig2}.
Therefore, we may expect that the fractal dimension of the particle E in three-dimensional Euclidean space lies in the range of $2.0 \le D \le 3.0$.

\citet{fulle-et-al2016} claimed that dust particles detected by GIADA (Grain Impact Analyzer and Dust Accumulator) on board Rosetta are extremely fluffy and characterized by $D = 1.87$\footnote{This value was evaluated for showers of fluffy dust particles detected only by the GDS (Grain Detection System) sub-system of GIADA.} consistent with $d_2 \approx 1.70 \pm 0.10$ for the particle E of MIDAS.
We should emphasize that dust particles detected only by GIADA/GDS are not necessarily characterized by $D < 2$, because there is no strong argument for the fractal dimension of $D = 1.87$ in \citet{fulle-et-al2016}.
According to \citet{fulle-et-al2015}, \citet{fulle-et-al2016} assumed $N=10^6$ for GDS-only dust particles whose typical geometric cross-section $A$ amounts to $A=8\times{10}^{-8}~\mathrm{m}^2$.
However, we are aware that \citet{fulle-et-al2015} did not provide solid evidence for $N=10^6$ and their arguments seem to be also consistent with $D \approx 3$ on the assumption of $N \approx 6 \times {10}^{8}$ and $po \approx 0.85$.
They attributed the fragile nature of the so-called GDS-only showers to the fluffiness of aggregate particles, but instead one could attribute it to the carbonization of organic matter, as will be discussed in the next section.
Consequently, the detection of the GDS-only showers by GIADA does not conflict with COSIMA's data as well as our current understandings of coagulation in the solar nebula and processing in a dust mantle.

\begin{figure}
\epsscale{0.5}
\plotone{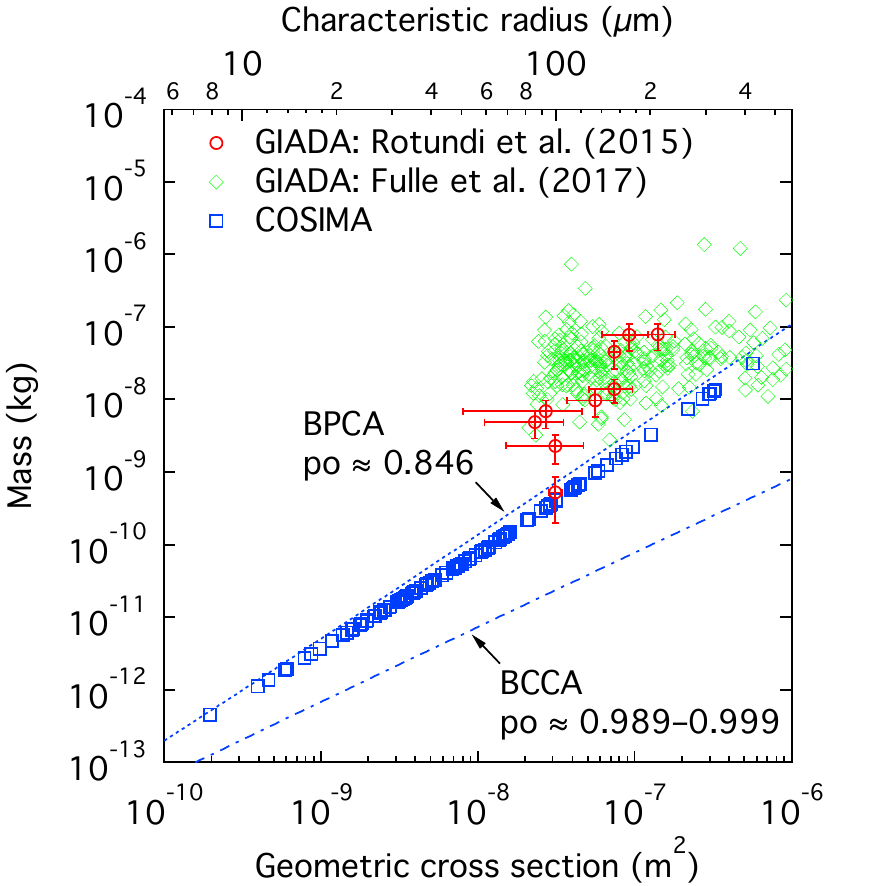}
\caption{The relation between the geometric cross section $A$ and the mass $m$ of dust particles detected by both the GDS and the IS sub-systems of GIADA \citep{rotundi-et-al2015,fulle-et-al2017}. 
The dash-dotted and dotted lines are the $A$-$m$ relations for ballistic cluster-cluster aggregates (BCCAs) and ballistic particle-cluster aggregates (BPCAs), respectively, formed under hit-and-stick coagulation process. 
Also plotted as squares are the $A$-$m$ relation derived from COSISCOPE images of dust particles collected by COSIMA.
The characteristic radius of a particle is intended to give an estimate of the particle size and here it is given by $({A/\pi})^{1/2}$ with $A$ being the geometric cross-section of the particle.
\label{fig3}}
\end{figure}

\citet{rotundi-et-al2015} and \citet{fulle-et-al2017} provide GIADA's data on the geometric cross section and the mass of each particle measured by both the GDS and the IS (Impact Sensor) sub-systems\footnote{While \citet{fulle-et-al2017} analyzed the whole set of GIADA's data inclusive of the limited data analyzed by \citet{rotundi-et-al2015}, it turned out that the $A$ and $m$ values derived from the same data set by \citet{fulle-et-al2017} and \citet{rotundi-et-al2015} do not coincide each other, implying that the values are strongly dependent on their calibrations.}.
Figure~\ref{fig3} compares the GIADA/GDS+IS data with the relationship between $A$ and $m$ for fractal aggregates grown under hit-and-stick coagulation process without restructuring of constituent grains (dotted line: BPCA particles; dash-dotted line: BCCA particles).
The $A$-$m$ relations for BCCA and BPCA particles formed under hit-and-stick coagulation processes without restructuring have been numerically determined as an analytical function of $N$ \citep{ossenkopf1993,minato-et-al2006}.
It turns out that the GIADA/GDS+IS data seem not to be inconsistent with the geometrical cross section $A$ of aggregate particles derived from their COSISCOPE images if the particles grow by the ballistic particle-cluster coagulation process (open squares). 
The slope of the GIADA's data are scattered around the $A$-$m$ relation for BPCA particles with $D \approx 3$ and $po \approx 0.85$, although the GIADA's data tend to exceed the mass of fractal aggregates with the same geometric cross section.
Because the slope of BCCA particles with $D \approx 2$ is much gentler compared to the one for higher fractal dimensions, the GIADA/GDS+IS data do not seem to be in harmony with fractal aggregates with $D < 3$.
The difference in the slope between GIADA's data and BPCA particles without restructuring might indicate that GIADA's aggregate particles suffered from restructuring either when they grew by coagulation under the ballistic particle-cluster process in the solar nebula or when they formed a dust mantle on the surface of comet 67P/Churyumov-Gerasimenko.

\subsection{Chemical Composition}

\begin{figure}
\epsscale{1.0}
\plotone{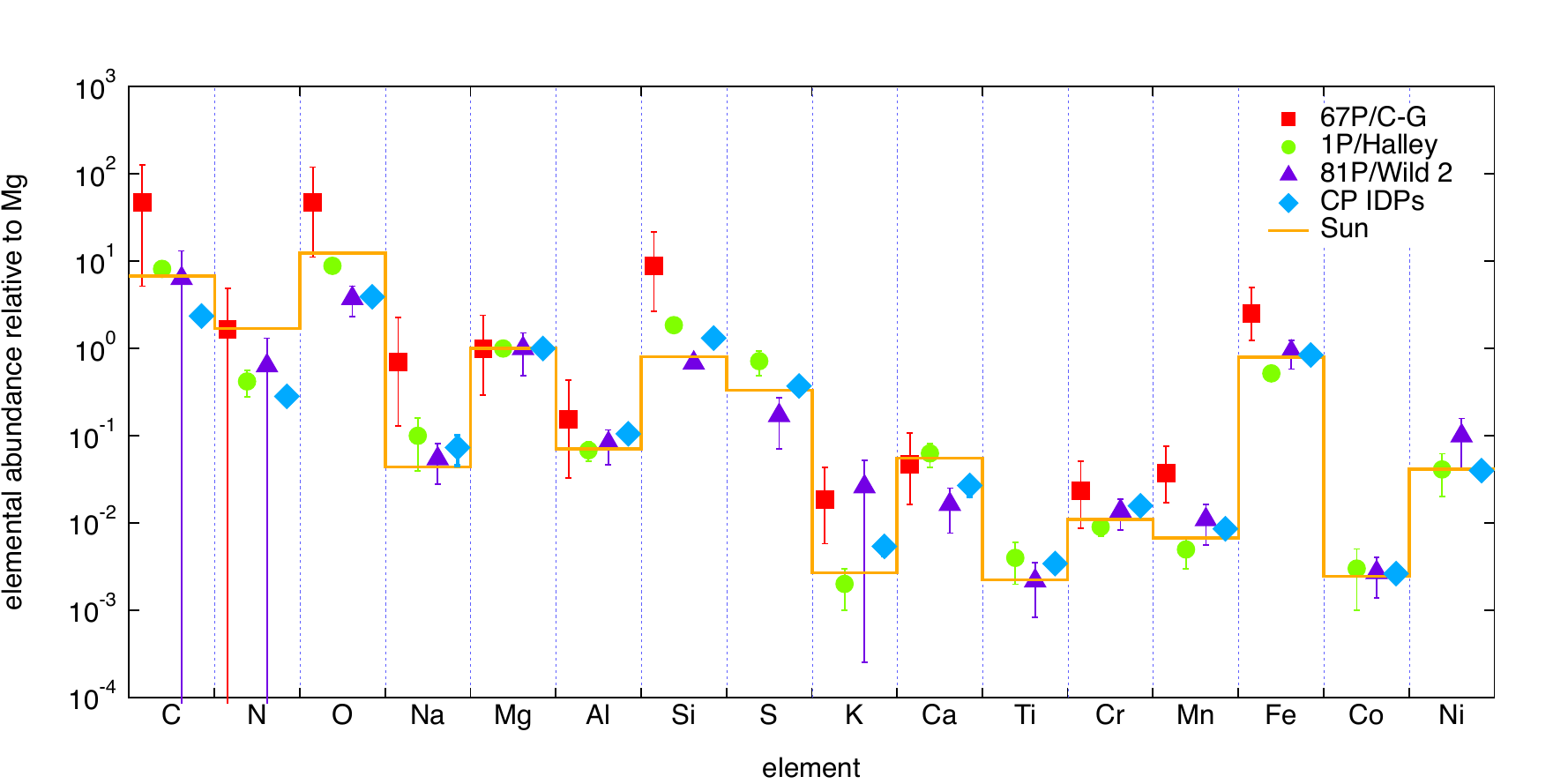}
\caption{The elemental abundances of dust particles in comet 67P/Churyumov-Gerasimenko normalized to the abundance of Mg (filled squares) measured by COSIMA \citep{bardyn-et-al2017}. 
Also plotted as filled circles, triangles, diamonds, and a solid line are the Mg-normalized elemental abundances of dust particles in comets 1P/Halley and 81P/Wild 2, CP IDPs, and the solar photosphere \citep{jessberger-et-al1988,stephan-et-al2008a,stephan-et-al2008b,leroux-et-al2008,cody-et-al2008,schramm-et-al1989,arndt-et-al1996,keller-et-al1995,asplund-et-al2009}.
\label{fig4}}
\end{figure}
The elemental abundances of dust particles in comet 67P/Churyumov-Gerasimenko have been derived from the mass spectra of 30 particles measured by COSIMA onboard Rosetta \citep{bardyn-et-al2017}.
Figure~\ref{fig4} compares the elemental abundances of dust particles in comet 67P/Churyumov-Gerasimenko with those in comets 1P/Halley derived from 79 high quality mass spectra measured by PUMA-1 onboard VeGa-1 and those in 81P/Wild 2 from laboratory analyses of cometary matter in aerogel samples and Al foil residues returned by Stardust \citep{jessberger-et-al1988,stephan-et-al2008a,stephan-et-al2008b,leroux-et-al2008,cody-et-al2008}.
We also plot the elemental abundances of CP IDPs collected in the stratosphere and the solar photosphere, by considering that CP IDPs are likely of cometary origin and dust particles in comets are originally condensates from the nebular gas of solar composition \citep{schramm-et-al1989,arndt-et-al1996,keller-et-al1995,asplund-et-al2009}.
Note that PUMA-1 measured the bulk composition of dust particles in comet 1P/Halley, most of which seem to consist of a chondritic core and an organic mantle \citep{kissel-krueger1987}.
In contrast, COSIMA is sensitive only to the surface matter of dust particles down to a depth of a few nanometers and thus the chemical composition of dust  aggregates might be biased toward surface materials of their constituent grains.
As a result, if chondritic materials are enclosed with carbonaceous matter, then COSIMA tends to trace the elemental abundances of the carbonaceous matter.

On the one hand, the elemental abundances of dust particles in comet 67P/Churyumov-Gerasimenko measured by COSIMA have revealed that Si is markedly supra-solar, when normalized to Mg.
On the other hand, one may realize that dust particles in comet 67P/Churyumov-Gerasimenko are depleted of Mg and Ca, when the elemental abundances are normalized to Si or Fe \citep[cf.][]{bardyn-et-al2017}.
Here we should emphasize that COSIMA does not necessarily probe the bulk composition of the chondritic materials, even if chondritic materials are not covered by carbonaceous matter.
A careful analysis of GEMS and pyroxene grains in CP IDPs has shown that the $\mathrm{Si/Mg}$ and $\mathrm{Si/Ca}$ atomic ratios are elevated near the surface of the grains, compared with the center of the grains \citep{bradley1994}.
By analogy with CP IDPs, the $\mathrm{Mg/Si}$ and $\mathrm{Ca/Si}$ ratios in comet 67P/Churyumov-Gerasimenko may be lower at the surfaces of constituent grains in dust aggregates than the centers of the grains (see Appendix~\ref{GEMS}). 
Therefore, we attribute the supra-solar $\mathrm{Si/Mg}$ ratio of dust particles in comet 67P/Churyumov-Gerasimenko to a systematic bias of COSIMA measurements in favor of surface materials.

The elemental abundances provide clues about the volume fractions of organic-rich carbonaceous matter and rock-forming components such as Mg-rich silicates, metals, and sulfides.
\citet{mann-et-al2004} have described how to derive the volume fractions of organic-rich carbonaceous matter and rock-forming components from elemental abundances.
They found that the volume fraction of carbonaceous matter, $f_\mathrm{CHON}$, and that of rock-forming matters, $f_\mathrm{ROCK}$, to be $f_\mathrm{CHON} = 0.66$ and $f_\mathrm{ROCK} = 0.34$ for dust particles in comet 1P/Halley \citep[cf.][]{kimura-et-al2006}.
In the same manner, we obtain $f_\mathrm{CHON} = 0.76$ and $f_\mathrm{ROCK} = 0.24$ for dust particles in comet 67P/Churyumov-Gerasimenko using the elemental abundances measured by COSIMA.
These estimates highlight the overall dominance of carbonaceous matter over rock-forming matters on the surfaces of dust particles from comets 67P/Churyumov-Gerasimenko and 1P/Halley.

We may estimate the mass fraction of carbonaceous matter, $\zeta_\mathrm{CHON}$, and that of rock-forming matters, $\zeta_\mathrm{ROCK}$, from their volume fractions, if their bulk densities are given. 
With a bulk density $\rho_\mathrm{CHON} = 1800~\mathrm{kg~m^{-3}}$ for carbonaceous matter and $\rho_\mathrm{ROCK} = 3500~\mathrm{kg~m^{-3}}$ for rock-forming matters, one gets $\zeta_\mathrm{CHON} \approx 0.5$ and $\zeta_\mathrm{ROCK} \approx 0.5$ for dust particles in the coma of comet 1P/Halley from their elemental abundances \citep{mann-et-al2004,kimura-et-al2006}.
\citet{fulle-et-al2018} obtained $\zeta_\mathrm{CHON} = 0.42 \pm 0.04$ and $\zeta_\mathrm{ROCK} = 0.58 \pm 0.04$ for dust particles in the coma of comet 67P/Churyumov-Gerasimenko from GIADA's data, by assuming a bulk density of $\rho_\mathrm{CHON} = 1200~\mathrm{kg~m^{-3}}$ for carbonaceous matter.
If we adopt $\rho_\mathrm{CHON} = 1800~\mathrm{kg~m^{-3}}$ instead, then we get $\zeta_\mathrm{CHON} = 0.48 \pm 0.07$ and $\zeta_\mathrm{ROCK} = 0.52 \pm 0.07$ for GIADA's data, but $\zeta_\mathrm{CHON} = 0.6$ and $\zeta_\mathrm{ROCK} = 0.4$ for COSIMA's data. 
The mass fractions derived from GIADA's data and COSIMA's data are marginally consistent, but the differences could be attributed to the fact that COSIMA tends to trace the carbonaceous matter, if it encloses rock-forming matter. 
As a result, there is no clear evidence that the mass fractions of carbonaceous matter and rock-forming matter in comet 67P/Churyumov-Gerasimenko greatly deviate from those in comet 1P/Halley.

\subsection{Responses to a Collision}

Hereafter we shall demonstrate how the outcome of collision for aggregate particles with the target of the COSIMA instrument depends on the composition of the particles.
For the sake of simplicity, we model cometary dust by aggregate particles consisting of $N$ identical spherical grains (monomers) of radius $r_0$ and volumetric mass density $\rho$.
The critical velocities for the onset of restructuring of an aggregate particle, $v_\mathrm{restr}$, the maximum compression of the aggregate, $v_\mathrm{comp}$, the onset of losing single spheres from the aggregate, $v_\mathrm{loss}$, and catastrophic disruption of the aggregate, $v_\mathrm{catastr}$, are given by \citep{chokshi-et-al1993,dominik-tielens1997,wada-et-al2007}\footnote{If an aggregate loses half the monomers or more upon collision with the target, the collisional outcome of the aggregate is referred to as catastrophic disruption.}
\begin{linenomath*}
\begin{eqnarray}
v_\mathrm{restr} &=& {\left({\frac{90 \pi \gamma \xi_\mathrm{crit}}{\rho r_0^2 N}}\right)}^{1/2} , \\
v_\mathrm{comp} &=& {\left({\frac{36 \pi \gamma \xi_\mathrm{crit}}{\rho r_0^2}}\right)}^{1/2} , \\
v_\mathrm{loss} &=& {\left[{\frac{62.37^3 \pi^2 \gamma^5 {\left({1-\nu^2}\right)}^2}{r_0^5 \rho^3 E^2}}\right]}^{1/6} , \\
v_\mathrm{catastr} &=& {\left[{\frac{207.9^3 \pi^2 \gamma^5 {\left({1-\nu^2}\right)}^2}{r_0^5 \rho^3 E^2}}\right]}^{1/6} ,
\end{eqnarray}
\end{linenomath*}
where $\gamma$, $E$, and $\nu$ are the surface energy, Young’s modulus, and Poisson’s ratio, respectively.
Here $\xi_\mathrm{crit}$ denotes the critical displacement that the contact area between the grains starts to move and we take $\xi_\mathrm{crit} = 0.2~\mathrm{nm}$ \citep{dominik-tielens1995}.
Note that the above formulae for critical velocities not only demonstrate the outcomes of numerical simulations based on the contact mechanics, but also agree with laboratory experiments of aggregate collisions \citep{kimura-et-al2015}\footnote{The rebound of aggregate particles sometimes observed in laboratory experiments has not well been modeled by numerical simulations on mutual collisions between aggregate particles \citep[cf.][]{wada-et-al2011}. It is, however, worthwhile noting that the rebound in the experiments by \citet{ellerbroek-et-al2017} seems to take place only for aggregate particles larger than $r_\mathrm{c} \ge 40~\micron$, which exceeds the largest size ($r_\mathrm{c} \sim 3~\micron$) in the simulations.}.
Table~\ref{tab:table} gives the physical properties of polyurethane (PUR), hydrogenated amorphous carbon, amorphous silica, and amorphous water ice, which are used to model organic matter, carbonaceous matter, silicate, and ice, respectively, in comets. 
\begin{deluxetable}{lCCCCCl}
\tablecaption{Physical properties of materials analogous to cometary matter \label{tab:table}}
\tablehead{
\colhead{Material} & \colhead{$\rho$} & \colhead{$\gamma$} & \colhead{$E$}  & \colhead{$\nu$}  & \colhead{$\epsilon$}  & \colhead{Reference}\\
\colhead{} & \colhead{(${10}^{3}~\mathrm{kg~m^{-3}}$)} & \colhead{($\mathrm{J~m^{-2}}$)} & \colhead{($\mathrm{GPa}$)} & \colhead{} & \colhead{} & \colhead{}
}
\startdata
%PUR & 1.238 & 0.07282697 & 1.15\times{10}^{-4} & 0.48 & 1.74 & 1, 2, 3\\
PUR & 1.2 & 0.073 & 1.2\times{10}^{-4} & 0.48 & 1.74 & 1, 2, 3\\
a-C:H & 1.7 & 0.034 & 120 & 0.30 & 5.86 & 3, 4, 5\\
a-SiO$_2$ & 2.0 & 0.24 & 70 & 0.17 & 3.63 & 6, 7, 8\\
a-H$_2$O & 1.0 & 0.11  & 7.0  & 0.25 & 107 & 7, 9, 10\\
\enddata
\tablecomments{(1) \citet{potschke-et-al2002}; (2) \citet{mcnicholas-rankilor1969}; (3) \citet{louh-et-al2005}; (4) \citet{piazza-morell2009}; (5) \citet{marques-et-al2003}; (6) \citet{kimura-et-al2015}; (7) This study; (8) \citet{henning-mutschke1997}; (9) \citet{wada-et-al2007}; (10) \citet{johari-whalley1981}.}
\end{deluxetable}

\begin{figure}
\epsscale{1.0}
\plottwo{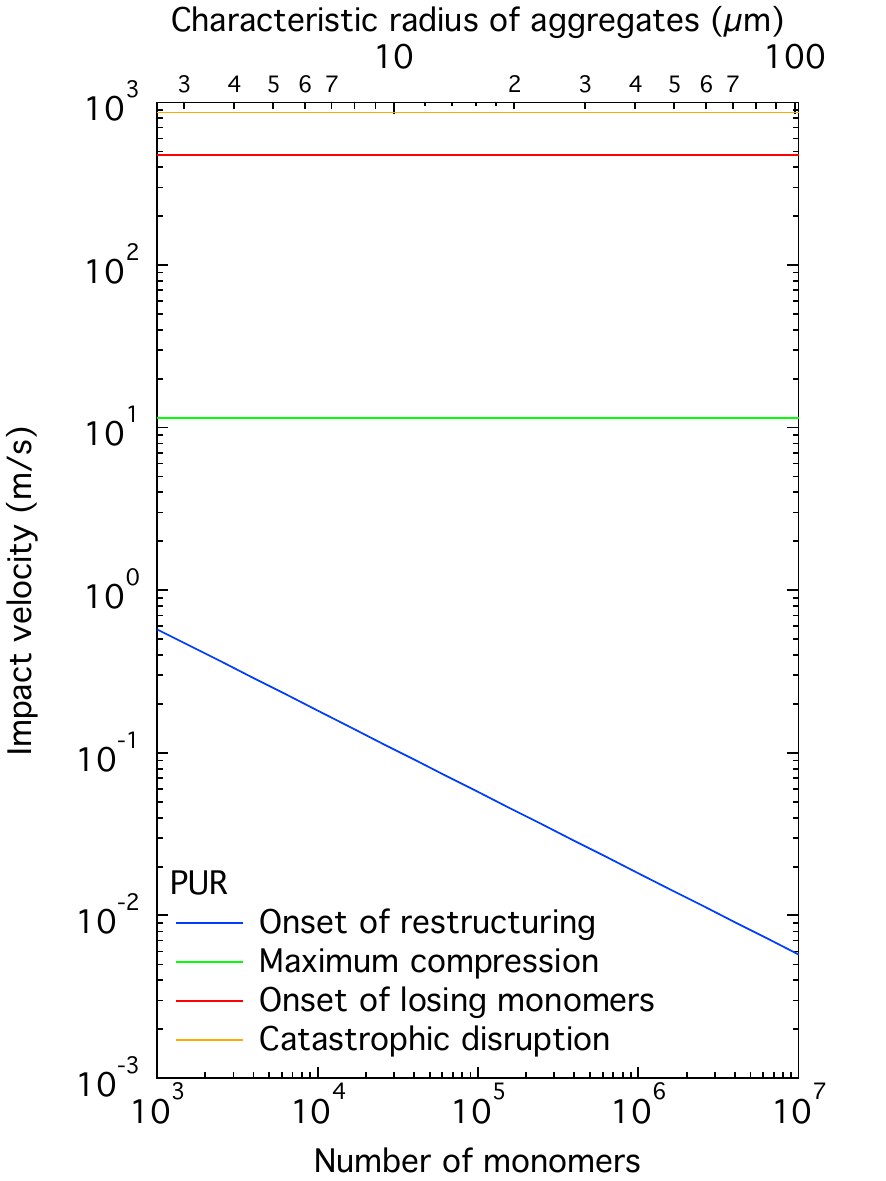}{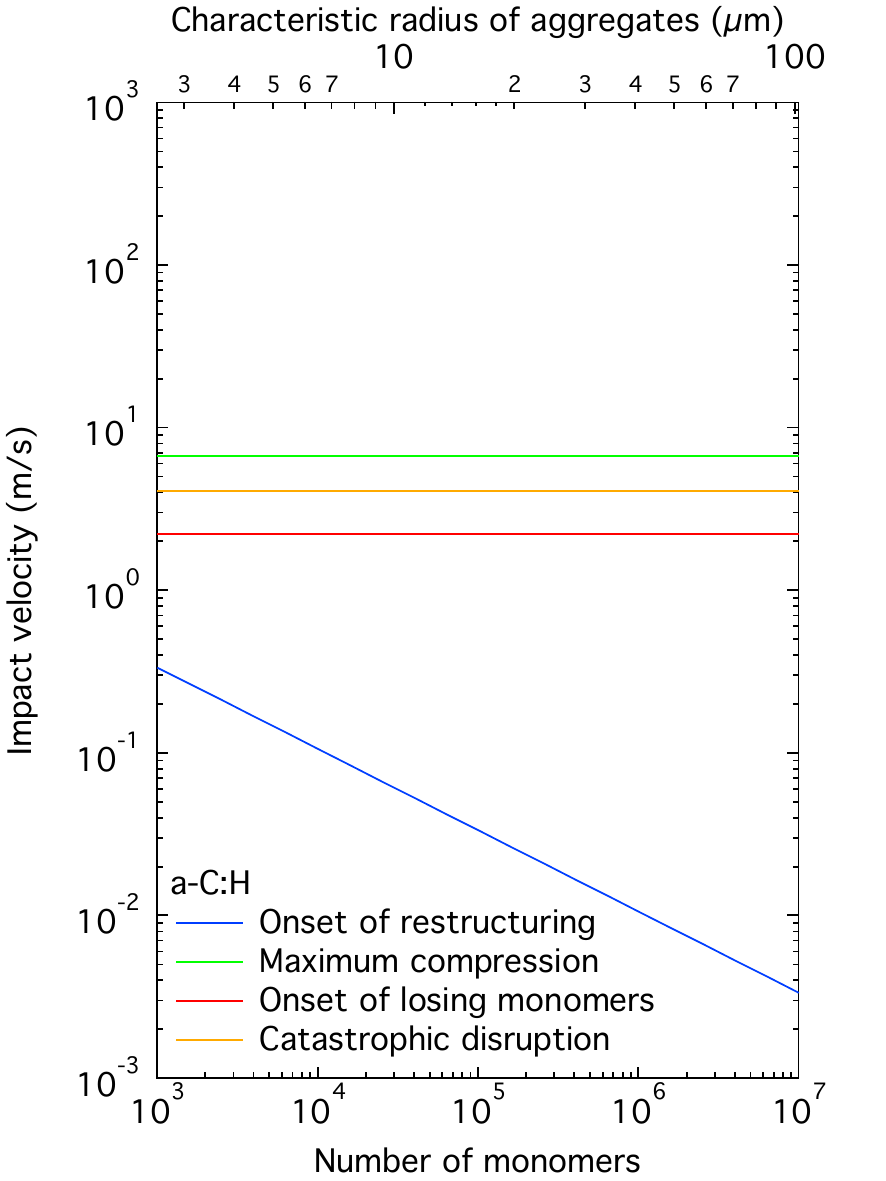}\\
\plottwo{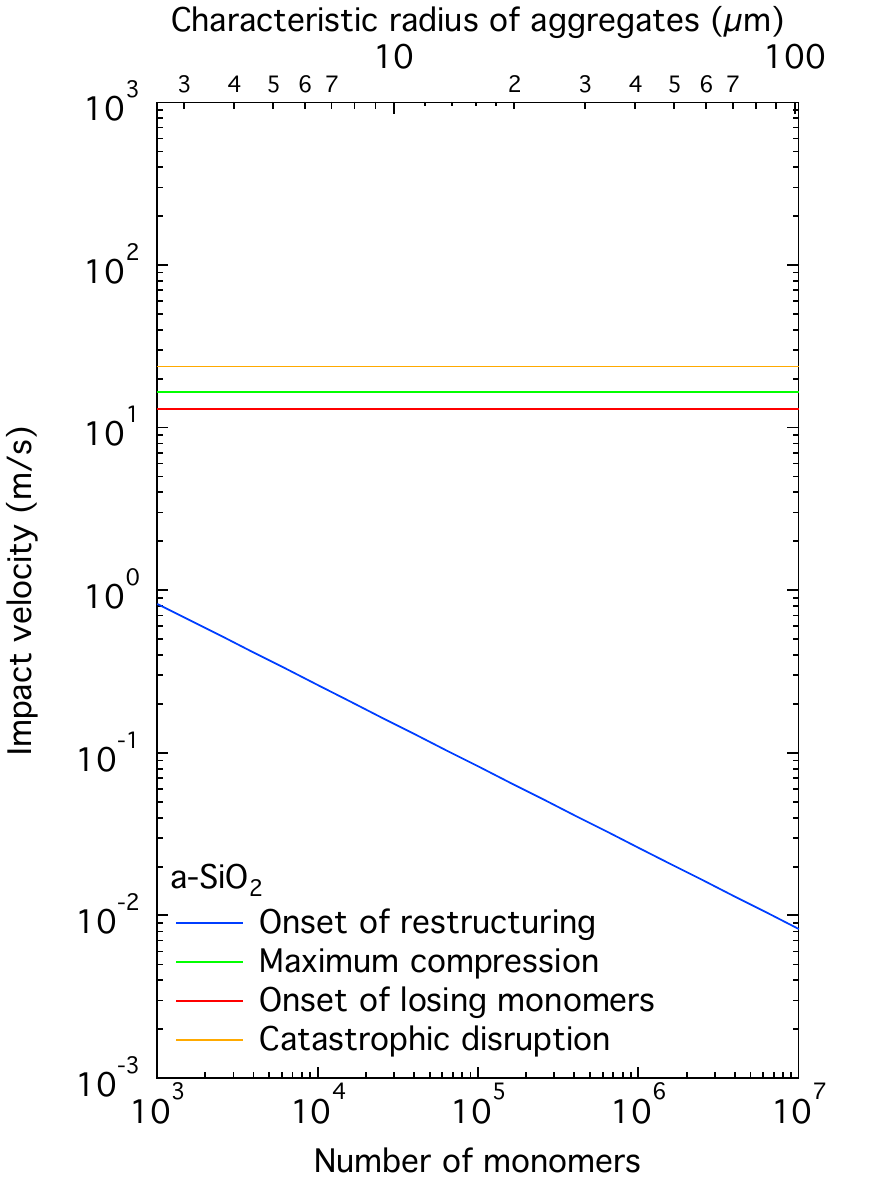}{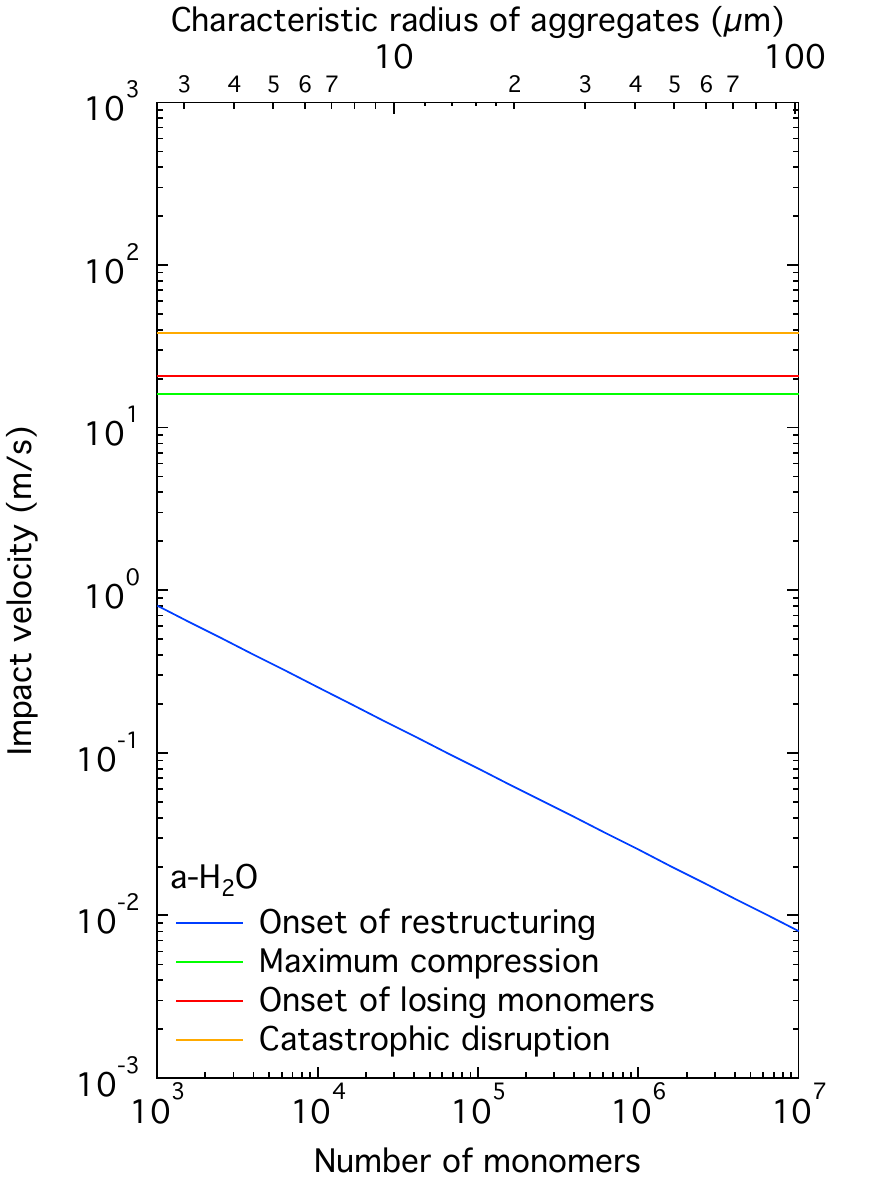}
\caption{The critical velocities for the onset of restructuring (blue), the maximum compression (green), the onset of losing monomers (red), and the catastrophic disruption (orange) of aggregate particles consisting of polyurethane (left top), hydrogenated amorphous carbon (right top), amorphous silica (left bottom), or amorphous water ice (right bottom) as a function of the number of monomers with radius $r_0 = 0.1~\micron$. 
The characteristic radius of the particles given along the horizontal top axis corresponds to that for fractal aggregates with a fractal dimension of $D=2.5$. \label{fig5}}
\end{figure}
Figure~\ref{fig5} shows the critical velocities $v_\mathrm{restr}$, $v_\mathrm{comp}$, $v_\mathrm{loss}$, and  $v_\mathrm{catastr}$ as a function of the number of constituent grains (monomers) in an aggregate particle consisting of polyurethane, hydrogenated amorphous carbon, amorphous silica, and amorphous water ice.
We find that aggregate particles do not experience any type of fragmentation at impact velocity below $475~\mathrm{m~s^{-1}}$ if they are composed of organic matter, but the carbonization of aggregate particles reduces the impact velocity of fragmentation.
It turns out that aggregate particles consisting of amorphous carbon grains with radius $r_0 = 0.1~\micron$ fragment if an impact velocity of the particles lies in the range of 2.2--$4.1~\mathrm{m~s^{-1}}$.
They are catastrophically disrupted at impact velocities above $4.1~\mathrm{m~s^{-1}}$ before they experience maximum compression, because of $v_\mathrm{comp} > v_\mathrm{catastr}$.
Aggregate particles of silicate or water ice grains with radius $r_0 = 0.1~\micron$ may suffer from restructuring, but not from fragmentation unless the impact velocity goes well beyond $10~\mathrm{m~s^{-1}}$.
COSIMA's microscope images have revealed that aggregate particles of apparent diameter larger than $100~\micron$ fragmented into small parts upon collision with the collection plates of the instrument \citep{hornung-et-al2016}.
The impact velocities of the aggregate particles onto the plates have been estimated to be $3.5 \pm 1.5~\mathrm{m~s^{-1}}$ \citep{rotundi-et-al2015}.
Therefore, our estimates on the outcome of collision support the idea that organic refractory component of cometary dust is carbonized during the formation of a dust mantle.
Moreover, Fig.~\ref{fig5} indicates that the fragments of aggregate particles found in the COSIMA instrument cannot be consistent with aggregate particles of submicrometer-sized grains whose surfaces are dominated by silicate nor water ice nor a mixture of both.
The carbon-rich nature of dust in the coma of comet 67P/Churyumov-Gerasimenko has been derived from the mass spectra of the dust by the COSIMA instrument \citep{bardyn-et-al2017}.
Therefore, we conclude that COSIMA's microscope images provide further evidence for the carbon-rich nature of the dust, indicating the carbonization of organic matter to a certain extent.

\subsection{Responses to an Electric Field}

\begin{figure}
\epsscale{1.0}
\plottwo{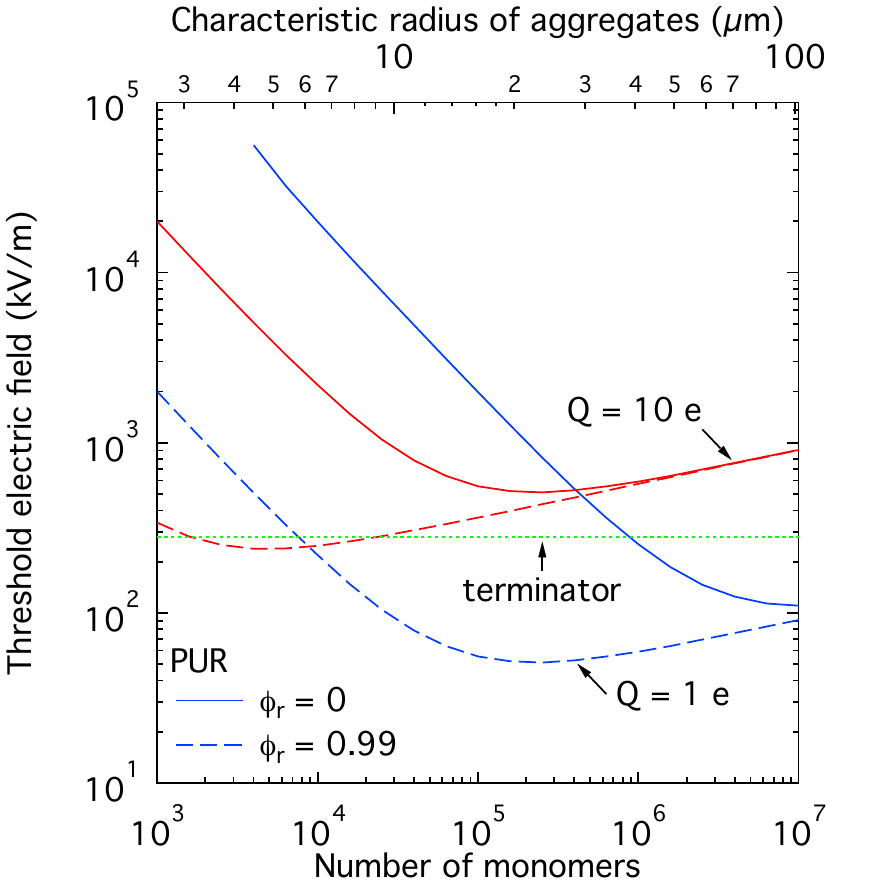}{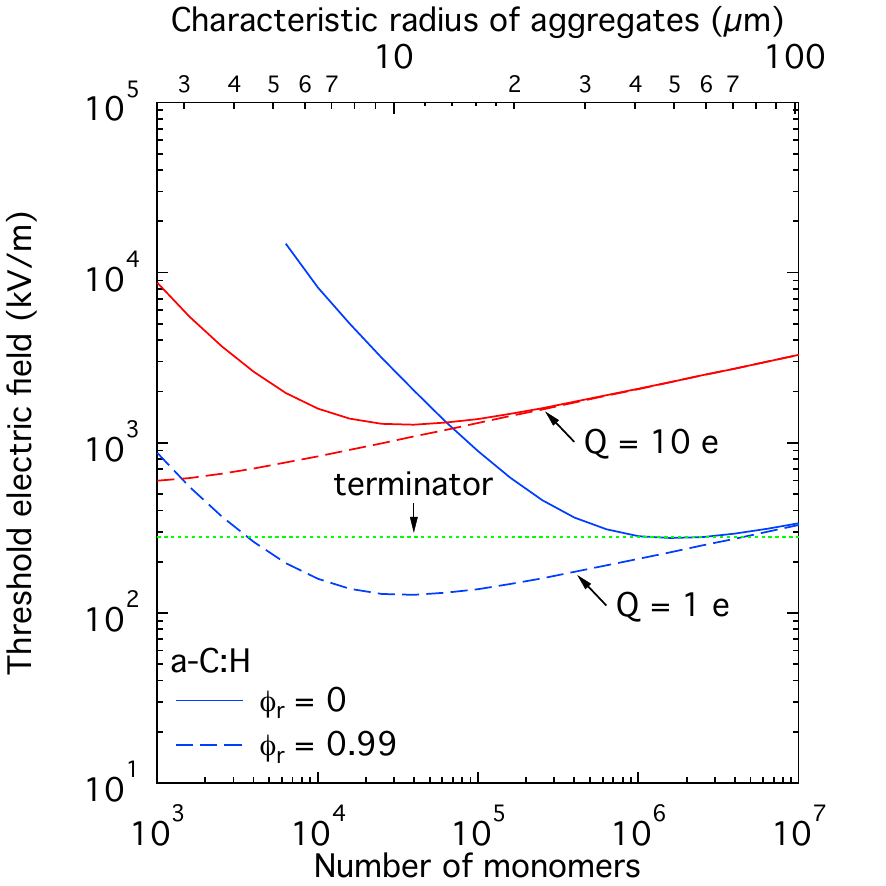}\\
\plottwo{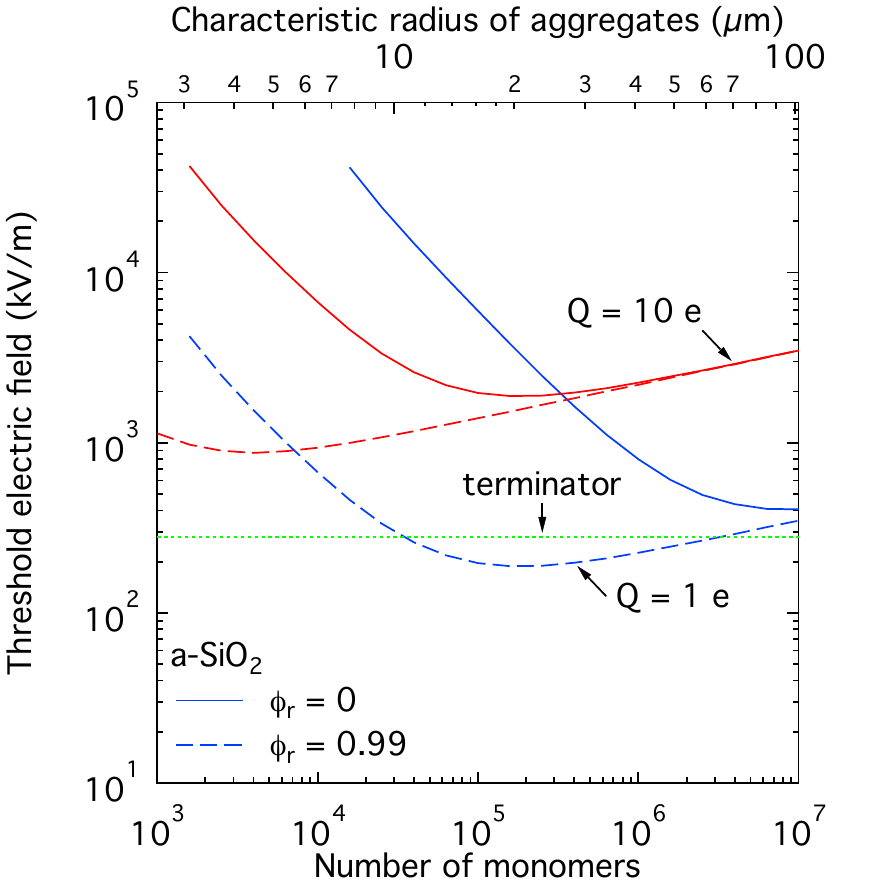}{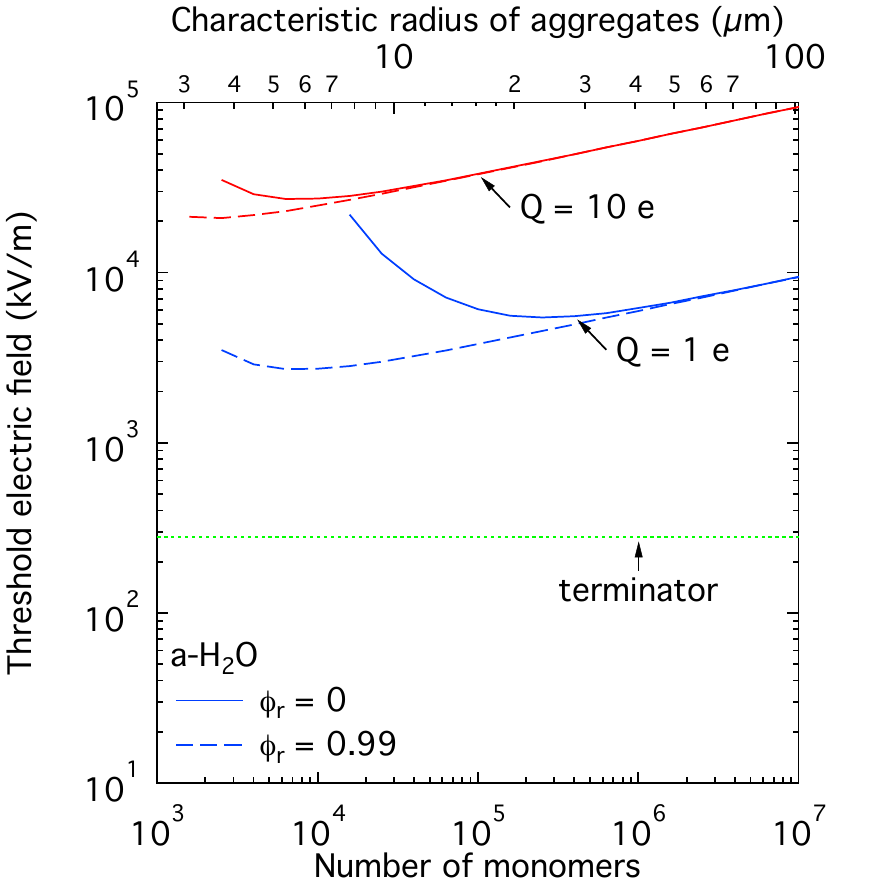}
\caption{The threshold electric field for lofting of aggregate particles consisting of dielectric grains as a function of the total number of grains (monomers) in an aggregate particle with fractal dimension $D = 2.5$. 
Left top: polyurethane; right top: hydrogenated amorphous carbon; left bottom: amorphous silica; right bottom: amorphous water ice. 
Here $(1- \phi_\mathrm{r})$ is a reduction factor of cohesive forces due to the roughness of grain surfaces (solid line: $\phi_\mathrm{r} = 0$; long dashed line: $\phi_\mathrm{r} = 0.99$) and $Q$ is the average electric charge per grain (red line: $Q = 10\,e$; blue line: $Q = 1\,e$). 
Also plotted as a dotted green line is the maximum strength of an electric field that could be achieved at the terminator. \label{fig6}}
\end{figure}

If an electric field is applied to aggregate particles, then the particles might be separated into two or more subunits, depending on the strength of the applied field and the composition of the particles \citep{kimura-et-al2014}.
By analogy with the Moon, \citet{mendis-et-al1981} suggested that lofting and hovering of dust particles take place on the surfaces of comets when outgassing activities of comet nuclei cease at heliocentric distances greater than 5~au from the Sun.
Dust aggregates in comets consist of submicrometer-sized grains whose surfaces are usually dominated by organic-rich carbonaceous material ($\sim 50~\mathrm{wt\%}$), in contrast to lunar agglutinates and Itokawa's samples \citep{greenberg1982,kissel-krueger1987,kimura-et-al2003}.
This is also the case for CP IDPs of supposedly cometary origin where submicrometer-sized GEMS grains are typically encased in organic-rich carbonaceous matter \citep{keller-et-al2000}.
The ratio of organic to silicate components varies among the grains and the aggregates, but no single grain was found to consist of pure silicate nor organic components in dust from comet 1P/Halley \citep{jessberger-et-al1988}.
It is worthwhile noting that a force acting on conductive particles in an electric field is repulsive, while a force acting on dielectric particles in an electric field is attractive \citep{arp-mason1977,nakajima-matsuyama2002}.
This dependence of electrostatic force on the grain material was confirmed by laboratory experiments on granular materials of 100~$\mu$m sized grains \citep{holstein-hathlou2012}.
Therefore, all forces acting on dust grains on the surfaces of airless bodies become attraction, if the outer layers of dust grains are covered by organic substrate, which is the most insulating material \citep[cf.][]{mccarty-whitesides2008}.
Nevertheless, electric charges on the surface of dust particles in an electric field exert additional forces on the particles and thus might be able to lift up or disrupt the particles.
The electrostatic repulsive force $F_\mathrm{el}$ on aggregate particles consisting of $N$ dielectric spheres with radius $r_0$ in an applied field $E$ is given by \citep[cf.][]{nakajima-sato1999,sow-et-al2013}
\begin{linenomath*}
\begin{eqnarray}
F_\mathrm{el} = - \alpha\frac{1}{16 \pi \varepsilon_0} \frac{N^2 Q^2}{r_\mathrm{c}^2} + \left[{1+\frac{1}{2}\left({\frac{\epsilon-1}{\epsilon+2}}\right)}\right] N Q E - \beta \frac{3}{2}\pi \varepsilon_0 r_0^2 \left({\frac{\epsilon-1}{\epsilon+2}}\right)^2 E^2 ,
\label{el_force}
\end{eqnarray}
\end{linenomath*}
where $Q$ is the average electric charge on each grain, $\epsilon$ is the static dielectric constant, and $\varepsilon_0$ is the permittivity of free space.
Note that only the second term inside the braces of Eq.~(\ref{el_force}) corresponds to the repulsive force due to the electric field and the first and the third terms are attractive forces due to image charges and induced dipoles, respectively.
The electrostatic repulsive force must overcome the adhesive force given by
\begin{linenomath*}
\begin{eqnarray}
F_\mathrm{co} = \frac{3}{2} \pi \gamma r_0 \left({1-\phi_\mathrm{r}}\right) n_\mathrm{c} ,
\label{co_force}
\end{eqnarray}
\end{linenomath*}
where $\left({1-\phi_\mathrm{r}}\right)$ is defined as the ratio of cohesive forces for a rough surface to a smooth one and  $n_\mathrm{c}$ is the number of contacts for the grain at the point of disruption.
We find that numerical results of $\alpha$ for a charged dielectric sphere on a dielectric flat surface by \citet{nakajima-sato1999} are reproduced to within approximately 20\% by $\alpha \approx 0.15 \epsilon^{1.2}$.
As far as aggregates consisting of identical dielectric spheres of $N \ga {10}^{2}$ are concerned, we may assume $\beta \approx {10}^{2}$ \citep{nakajima-matsuyama2002}.
The threshold of electric field strength $E_\mathrm{th}$, below which dust particles cannot be detached from the surface by electric field, can be determined by the balance between Eqs.~(\ref{el_force}) and (\ref{co_force}):
\begin{linenomath*}
\begin{eqnarray}
E_\mathrm{th} & = & \frac{2N \left({\epsilon+2}\right) \left({\epsilon+1}\right)}{\beta \varepsilon_0 \left({\epsilon-1}\right)^{2}} \left({\frac{Q}{4 \pi r_0^2 }}\right) \left\{{1 - \sqrt{1 - \beta \left({\frac{\epsilon-1}{\epsilon+1}}\right)^2 \left[{\frac{\alpha}{6}\left({\frac{k_0}{N}}\right)^{2/D} + \frac{\varepsilon_0 \gamma \left({1-\phi_\mathrm{r}}\right) n_\mathrm{c}}{4 r_0 N^2} \left({\frac{Q}{4 \pi r_0^2}}\right)^{-2}}\right] }}\right\} .
\label{e_th_insulator}
\end{eqnarray}
\end{linenomath*}
Note that $E_\mathrm{th}$ in Eq.~(\ref{e_th_insulator}) may be underestimated, because the gravitational attractive force acting on the particles from the surface must elevate the value of $E_\mathrm{th}$.
Figure~\ref{fig6} shows the dependence of $E_\mathrm{th}$ on the number $N$ of dielectric spheres in an aggregate particle with a fractal dimension of $D = 2.5$.
The particles are assumed to consist of organic matter (left top), hydrogenated amorphous carbon (right top), amorphous silica (left bottom), or amorphous water ice (right bottom).
If the outer layer of cometary dust is covered by water ice and contains one elementary charge per grain on average or more, we expect that the dust cannot be lofted off the surfaces of short-period comets by electrostatic forces even at the terminator nor does hovering of grains take place.

\begin{figure}
\epsscale{1.0}
\plottwo{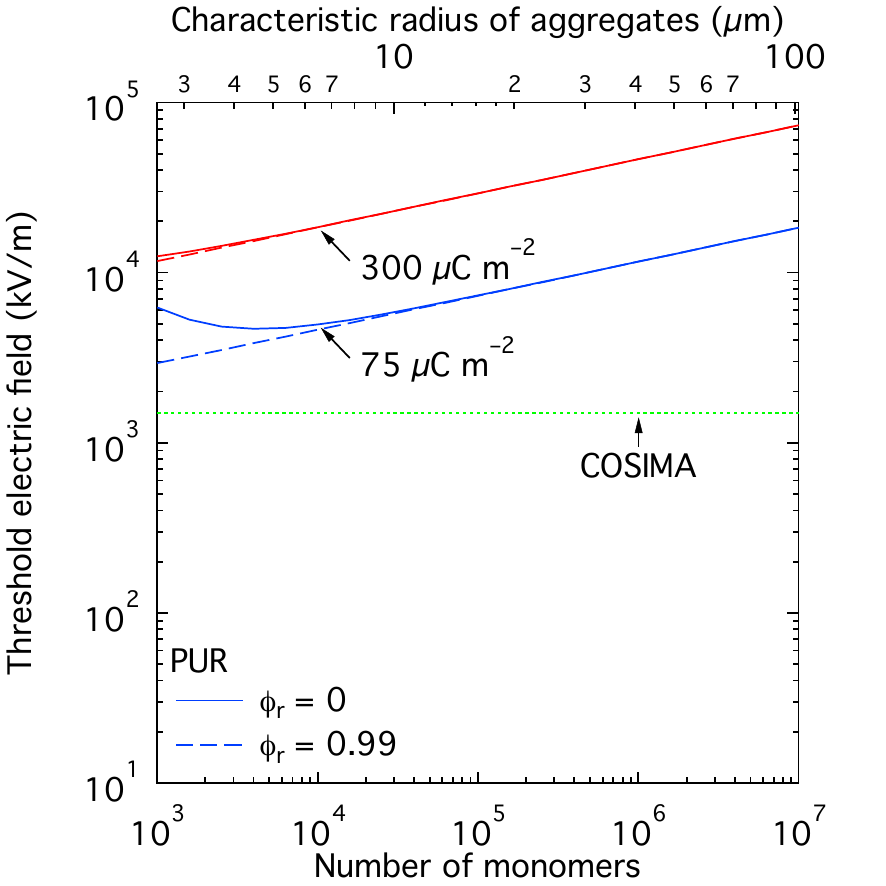}{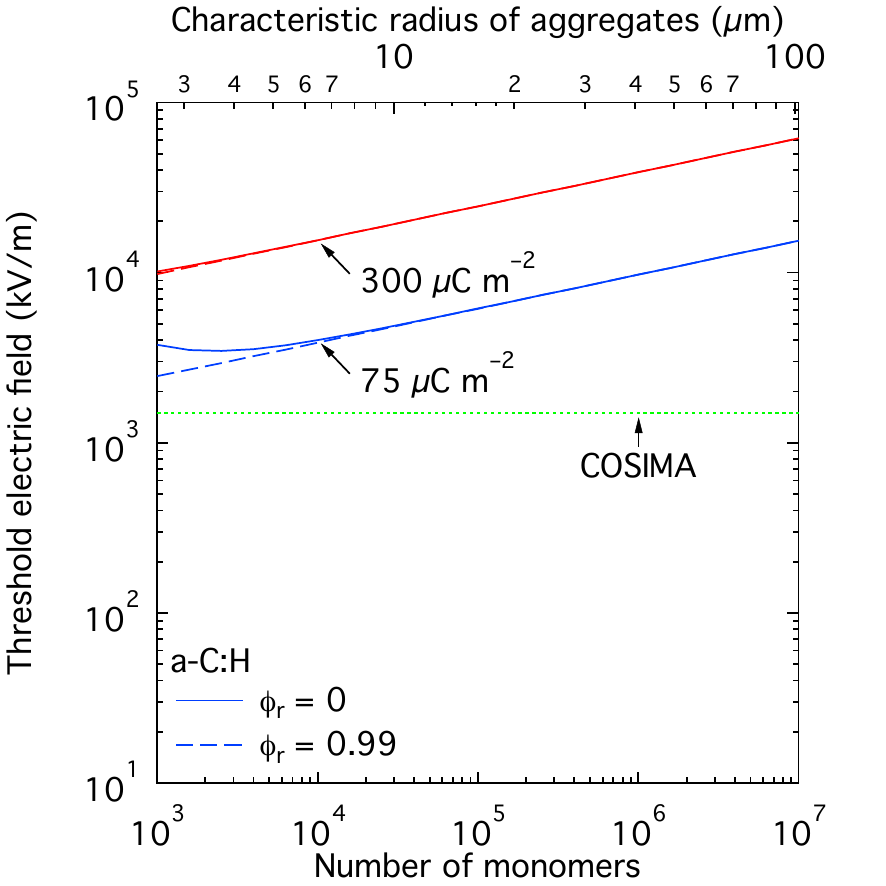}
\caption{The same as Fig.~\ref{fig6}, but in the case for the surface charge densities of $75~\mathrm{\mu C~m^{-2}}$ (blue line) and $300~\mathrm{\mu C~m^{-2}}$ (red line) on the surfaces of polyurethane (left) and hydrogenated amorphous carbon (right).
Also plotted as a dotted line is the strength of an electric field that was applied to cometary dust by COSIMA's experiments. \label{fig7}}
\end{figure}

COSIMA's experiments on collected particles have observed fracture of an aggregate particle in an electric field of $1.5~\mathrm{MV~m^{-1}}$ after sufficient exposure to indium ion beams \citep{hilchenbach-et-al2017}.
The experimental result was interpreted as evidence for accumulation of positive charges on the particle on the order of 75--$300~\mathrm{\mu C~m^{-2}}$ by sequential irradiation of indium ion beams.
To investigate whether the electrostatic repulsive force on insulators in the COSIMA's electric field is strong enough to overcome the cohesive force, we shall estimate the minimum strength $E_\mathrm{th}$ of electric field for the electrostatic lofting of aggregate particles. 
Because COSIMA collected dust particles by their impacts on metal targets, we may assume $\alpha = 1$ in the first term of Eq.~(\ref{el_force}).
Figure~\ref{fig7} shows the dependence of the threshold electric field $E_\mathrm{th}$ on the number $N$ of dielectric spheres in an aggregate particle with $D=2.5$.
It turns out that the results are insensitive to the choices of $\beta$, $\gamma$, and $\phi_\mathrm{r}$ because the image charge force (the first term of Eq.~(\ref{el_force})) is the most dominant attractive force.
If positive charges on the order of 75--$300~\mathrm{\mu C~m^{-2}}$ are accumulated on the surface of aggregate particles, then the attractive force is too large to lift up the particles from the metal targets.

\begin{figure}
\epsscale{0.5}
\plotone{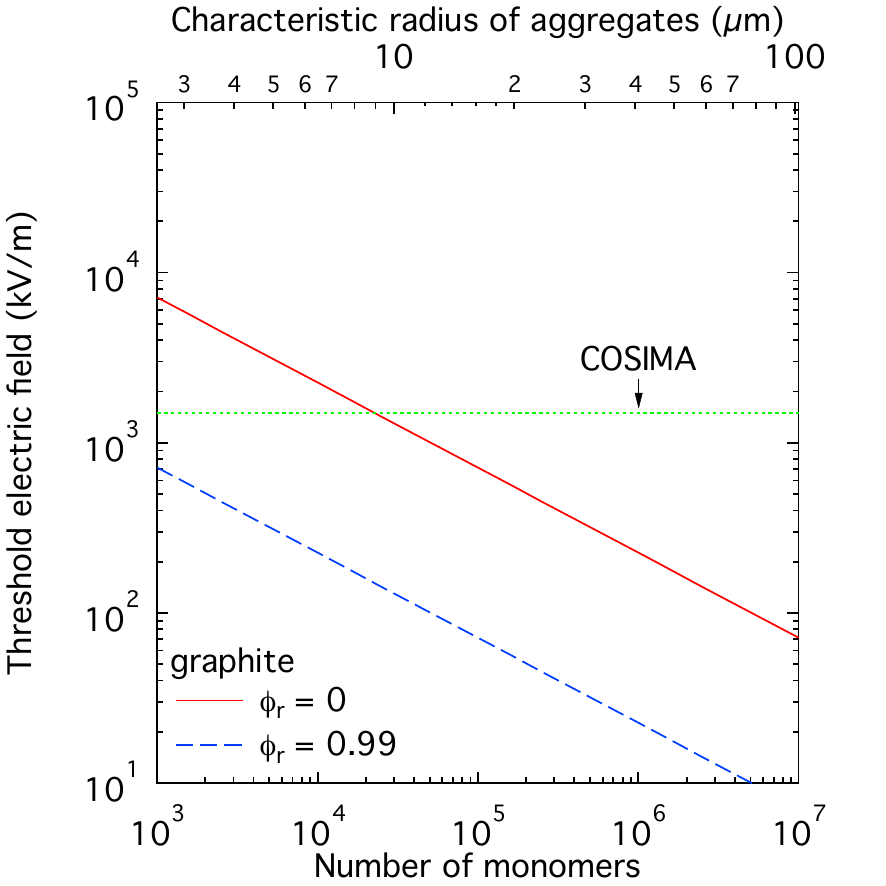}
\caption{The threshold electric field for lofting of aggregate particles consisting of graphite (representing amorphous carbon with zero band gap) grains with surface roughness of $\phi_\mathrm{r} = 0$ (solid line) and $\phi_\mathrm{r} = 0.99$ (long dashed line) as a function of the total number of grains (monomers) in an aggregate particle with fractal dimension $D = 2.5$. 
Also plotted as a dotted line is the strength of an electric field that was applied to cometary dust by COSIMA's experiments.  \label{fig8}}
\end{figure}

Photo-processing of carbonaceous material due to exposure to ultraviolet (UV) radiation reduces the band gap and enhances the $\mathrm{sp}^2/\mathrm{sp}^3$ ratio \citep{jones2012}.
Consequently, dielectric properties may be transformed into semimetallic properties by graphitization, as the carbonization of organic matter proceeds with exposure to UV radiation.
Therefore, an external electric field might be able to lift up aggregate particles of highly carbonized grains, but not aggregate particles of less carbonized grains.
COSIMA's experiments on collected particles have shown electrostatic lofting of some particles when an electric field of $1.5~\mathrm{MV~m^{-1}}$ is applied to the particles \citep{hilchenbach-et-al2017}.
COSIMA's results may indicate that the lofted particles were highly carbonized by solar UV radiation on the surface of the comet or even graphitized to some extent. 
To investigate whether the electrostatic repulsive force on conductors in the COSIMA's electric field is strong enough to overcomes the cohesive force, we shall estimate the minimum strength $E_\mathrm{th}$ of electric field for the electrostatic lofting of conducting aggregate particles. 
If the aggregate particles consist of $N$ spherical conducting monomers having radius $r_0$, then $E_\mathrm{th}$ is given by \citep{kimura-et-al2014} 
\begin{linenomath*}
\begin{eqnarray}
E_\mathrm{th} & = & \min \left[{\frac{\pi^3 r_0 \gamma \left({1-\phi_\mathrm{r}}\right) n_\mathrm{c}}{4e \left({\zeta(3)+\frac{1}{6}}\right)} , \; \sqrt{\frac{3\gamma \left({1-\phi_\mathrm{r}}\right)}{8\varepsilon_0 \left({\zeta(3)+\frac{1}{6}}\right)r_0} \left({\frac{n_\mathrm{c}}{N}}\right)}}\right],
\label{e_th_conductor}
\end{eqnarray}
\end{linenomath*}
where $e$ is the elementary charge and $\zeta$ is Riemann's zeta function.
Figure~\ref{fig8} depicts the dependence of $E_\mathrm{th}$ on the size of the particles, on the assumption that the surface energy of amorphous carbon with zero band gap is equivalent to that of graphite (i.e., $\gamma = 0.165~\mathrm{J~m^{-2}}$) \citep[cf.][]{abrahamson1973}.
We find that the strength of electric field of $1.5~\mathrm{MV~m^{-1}}$ is sufficient to loft aggregates with characteristic radius $r_\mathrm{c} \ga 9~\micron$.
However, it should be noted that $E_\mathrm{th}$ in Eq.~(\ref{e_th_conductor}) might be overestimated by more than an order of magnitude, because the rugosity of each monomer enhances the induction charge and reduces the cohesion.
Therefore, we expect that the electrostatic lofting of aggregate particles could be observed in COSISCOPE even for the smallest size of the particles, if their surfaces are heavily carbonized.
This may, however, not be the case for most of the particles, because the electrostatic lofting of aggregate particles in COSIMA's experiments was assisted with indium ion beams. 
 
\begin{figure}
\epsscale{0.5}
\plotone{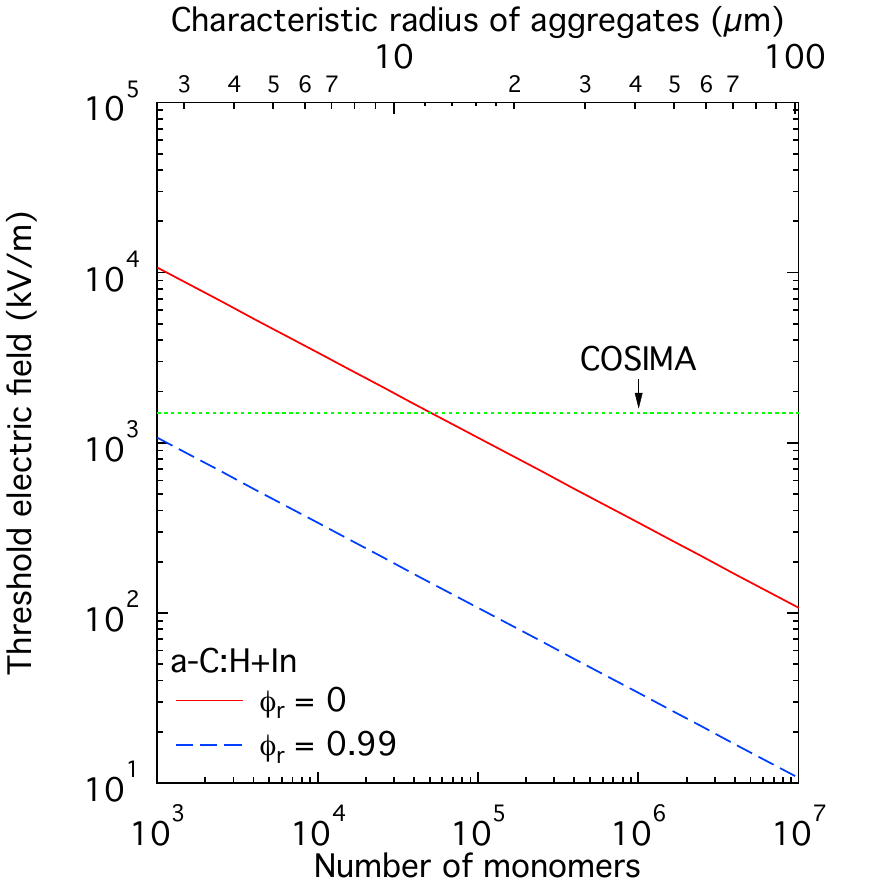}
\caption{The same as Fig.~\ref{fig8}, but for hydrogenated amorphous carbon covered with a thin layer of indium. \label{fig9}}
\end{figure}

Because the molecular structure of a surface is easily modified by adsorption of atoms and molecules on the surface, an insulating surface could behave like a conducting one, depending on the adsorbed atoms and molecules on the surface \citep{voorthuyzen-et-al1987,kimura-et-al2014}.
This is also the case for implantation of indium ions on the surface of an insulator that results in the formation of a thin metallic film on the surface of the insulator \citep{yoshimura-et-al2010}.
Therefore, it is most likely that once a dose of indium ion radiation exceeds a threshold, then even aggregate particles of less carbonized grains could behave like conductors and be lofted off by an applied electric field.
In Figure~\ref{fig9}, we plot the size dependence of $E_\mathrm{th}$ for aggregate particles of hydrogenated amorphous carbon grains with indium coating, using $\gamma = 0.37~\mathrm{J~m^{-2}}$ (see Appendix~\ref{adsorption}).
It turns out that $E_\mathrm{th}$ is low enough for aggregate particles of hydrogenated amorphous carbon grains with indium coating to be lofted off the COSIMA's target by an electric field applied in COSIMA.
Therefore, we expect that the electrostatic lofting of aggregate particles could be observed in COSISCOPE, irrespective of their size, if their surfaces are exposed to sufficiently high doses of indium ions.

\begin{figure}
\epsscale{0.5}
\plotone{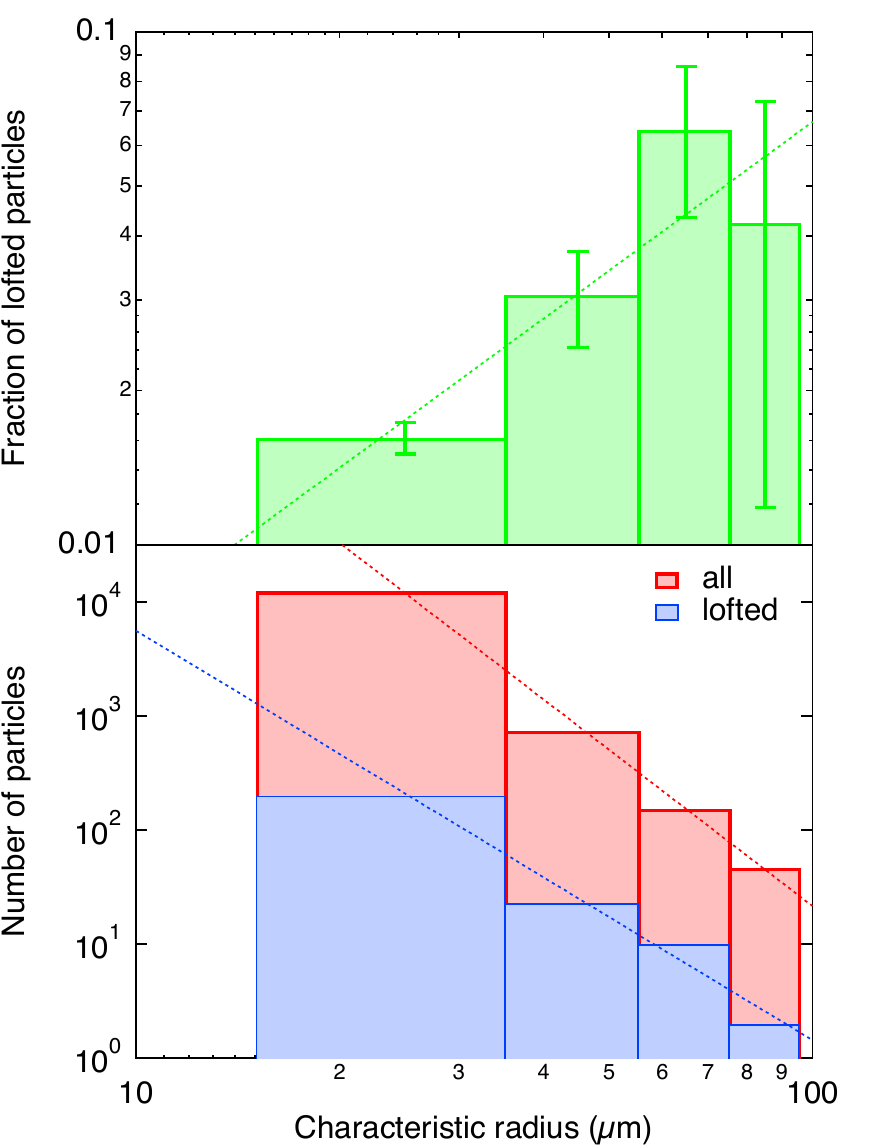}
\caption{The size distribution of dust particles derived from COSISCOPE images (red) and the number of the particles lofted off the target surface in an applied electric field (blue) in the lower panel, and the fraction of the lofted particles (green) in the upper panel. \label{fig10}}
\end{figure}

Eq.~(\ref{e_th_conductor}) indicates that the larger the size of aggregate particles is, the easier the electrostatic lofting of the particles is, in the case of conducting monomers.
Figure~\ref{fig10} depicts the size distribution of dust particles derived from COSISCOPE images, the number of the particles lofted off the target surface in an applied electric field, and the fraction of the lofted particles in each size bin.
Dotted lines show fitting lines in logarithmic scales and the error bars for the fraction of lofted particles are computed from square roots of the numbers for all particles and lofted particles.
The fraction of lofted particles tend to increase with the size of the particles, in accordance with responses of electrically conductive particles to an electric field given in Eq.~(\ref{e_th_conductor}).
This proves that the surface layer of dust particles imaged by COSISCOPE is at least in part composed of conductive material rather than dielectric material.

\section{Summary} 

We have examined the morphological, elastic, and electric properties of cometary dust using optical microscopic images of the dust taken by COSISCOPE of the COSIMA instrument onboard Rosetta.
Cometary dust is an aggregate of subunits, the spatial arrangement of which is characterized by the concept of fractals and consistent with the so-called rainout growth of $10~\micron$-sized particles in the solar nebula.
Electric responses of aggregates to an electric field in COSIMA indicate that the surfaces of the aggregates are dominated by dielectric materials, while their elastic responses to a collision onto the COSIMA target favor the surfaces covered by carbonaceous matter rather than silicates and ices.
Since the subunits of aggregates were most likely glued together by organic matter during the coagulation growth of aggregates in the solar nebula, COSIMA's experiments imply that organic matter was carbonized during the formation of a dust mantle.
Consequently, COSISCOPE images of cometary dust have embodied the most plausible scenario for the formation and evolution of comets, in which solar nebula condensates conglomerated into comets and solar radiation alters the surface structure and composition of comets in the inner solar system.

%% If you wish to include an acknowledgments section in your paper,
%% separate it off from the body of the text using the \acknowledgments
%% command.

\acknowledgments

We would like to thank Marco Fulle for useful discussion on the data for the cross section and mass of dust particles taken by GIADA onboard Rosetta.
We also thank both the reviewers for their careful readings and beneficial comments on the manuscript.
We acknowledge support from the Faculty of the European Space Astronomy Centre (ESAC), contract number PO 5001018751.
H.K. is grateful to a stipend from the Max Planck Institute for Solar System Research and JSPS's Grants-in-Aid for Scientific Research (KAKENHI \#23244027, \#26400230, \#15K05273, \#19H05085).

%% To help institutions obtain information on the effectiveness of their 
%% telescopes the AAS Journals has created a group of keywords for telescope 
%% facilities. 

%% Following the acknowledgments section, use the following syntax and the
%% \facility{} macro to list the keywords of facilities used in the research 
%% for the paper.  Each keyword is check against the master list during
%% copy editing.  Individual instruments can be provided in parentheses,
%% after the keyword, but they are not verified.

%\vspace{5mm}
%\facilities{HST(STIS), Swift(XRT and UVOT), AAVSO, CTIO:1.3m,
%CTIO:1.5m,CXO}

%\software{IRAF, cloudy, IDL}

%% Appendix material should be preceded with a single \appendix command.
%% There should be a \section command for each appendix. Mark appendix
%% subsections with the same markup you use in the main body of the paper.

%% Each Appendix (indicated with \section) will be lettered A, B, C, etc.
%% The equation counter will reset when it encounters the \appendix
%% command and will number appendix equations (A1), (A2), etc.

\clearpage

\appendix

\section{Depletion of Magnesium and Calcium From the Surface of CP IDPs}
\label{GEMS}

\begin{figure}
\epsscale{1.0}
\plotone{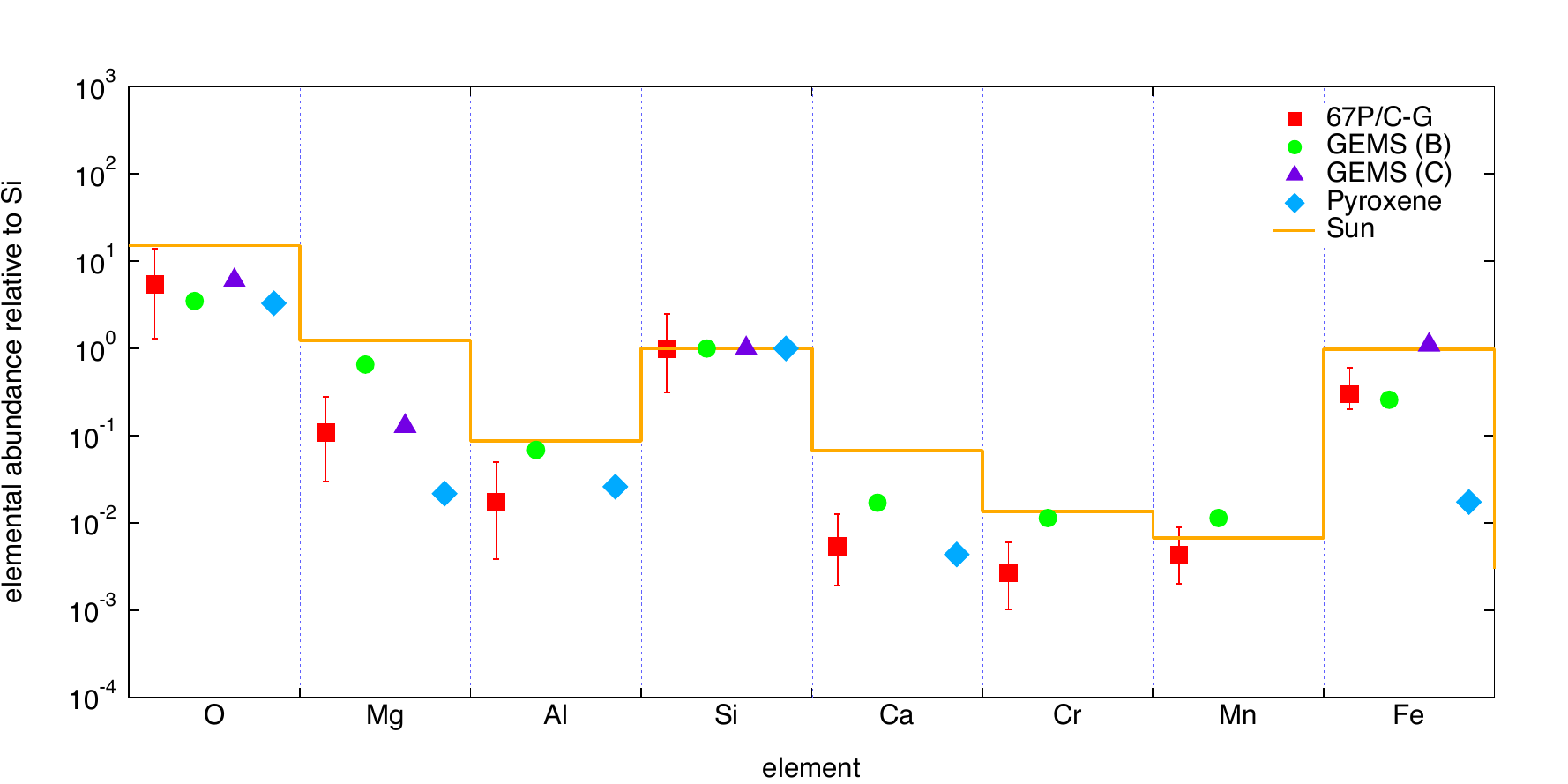}
\caption{The elemental abundances of dust particles in comet 67P/Churyumov-Gerasimenko normalized to the abundance of Si (filled squares) measured by COSIMA \citep{bardyn-et-al2017}.
Also plotted as filled circles, triangles, and diamonds are the Si-normalized elemental abundances of two GEMS grains and pyroxenes at their surfaces \citep{bradley1994}.
The solar photospheric abundances are depicted as a solid line, after they are normalized to Si \citep{asplund-et-al2009}.
\label{figa1}}
\end{figure}

The compositional profiles of two GEMS grains in CP IDPs revealed a gradual variation in the chemical composition from the core to the surface \citep{bradley1994}.
Figure~\ref{figa1} compares the abundances of major GEMS-forming elements between dust particles in comet 67P/Churyumov-Gerasimenko (filled squares) and the surfaces of two GEMS grains (filled circles and triangles).
Also plotted as filled diamonds and solid lines are the elemental abundances measured at the surface of a low-Fe pyroxene crystal in CP IDPs and the solar photospheric abundances, respectively \citep{bradley1994,asplund-et-al2009}.
If we take geometric means of the $\mathrm{Mg/Si}$ and $\mathrm{Ca/Si}$ atomic ratios over the GEMS and the pyroxene, we obtain $\mathrm{Mg/Si} = 0.12^{+0.55}_{-0.09}$ and $\mathrm{Ca/Si} = 0.0087^{+0.0141}_{-0.0054}$ for the mineral surfaces.
%Mg/Si=0.12252128174557494, Ca/Si=0.008652146449200652
Since the corresponding values of 67P/Churyumov-Gerasimenko are $\mathrm{Mg/Si} = 0.11^{+0.17}_{-0.08}$ and $\mathrm{Ca/Si} = 0.0053^{+0.0072}_{-0.0035}$, the depletions of Mg and Ca relative to Si found in comet 67P/Churyumov-Gerasimenko by COSIMA seem to be consistent with the chemical compositions of IDP's mineral surfaces.

\section{Estimates of Surface Energy From Surface Tension}
\label{surface-tension}

\begin{figure}
\epsscale{1.0}
\plottwo{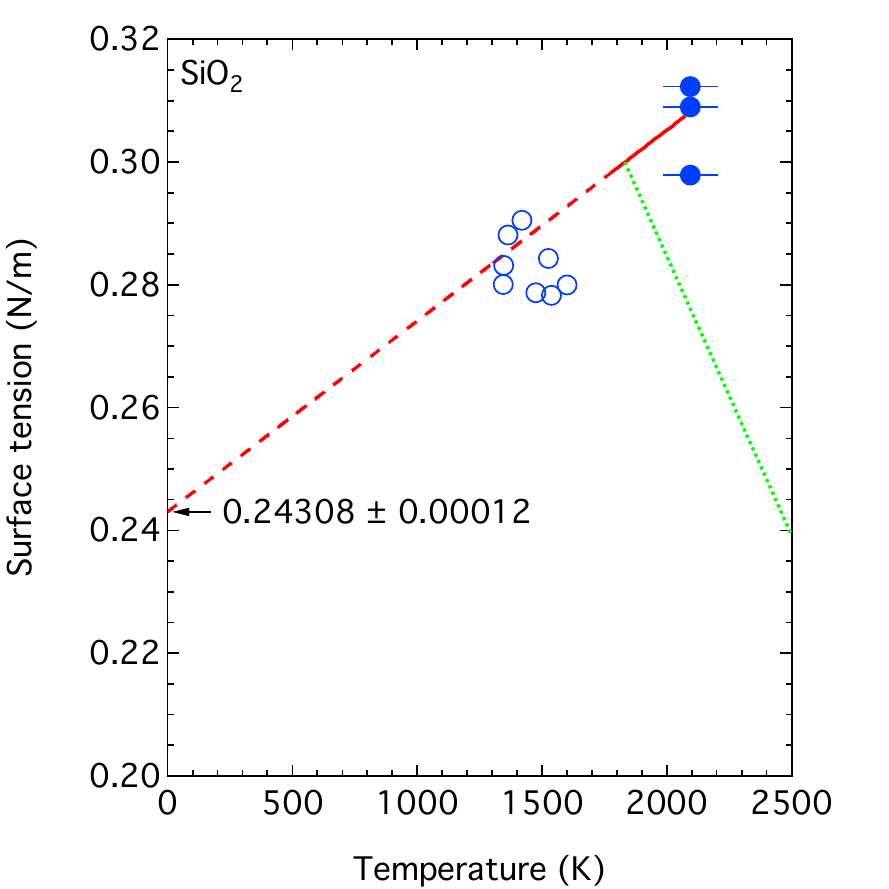}{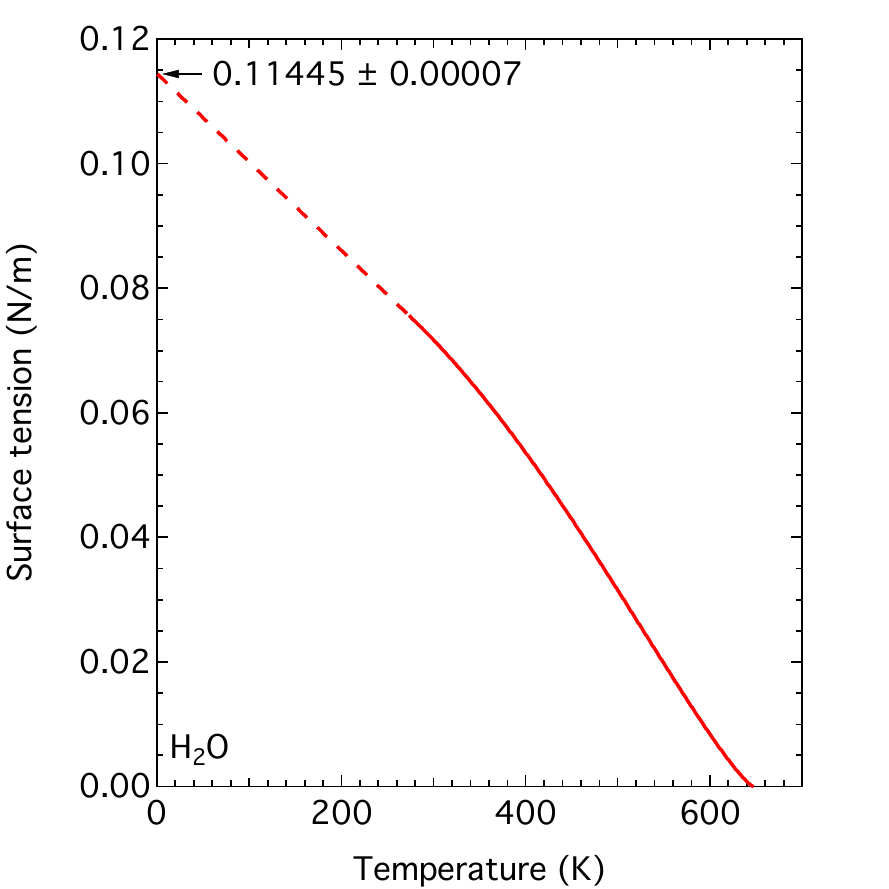}
\caption{The dependence of surface tension on the temperature for amorphous $\mathrm{SiO_2}$ (left) and $\mathrm{H_2O}$ (right). 
Open circles: experimental data by \citet{parikh1958}; closed circles: experimental data by \citet{kingery1959}; dotted line: a model by \citet{kraus-et-al2012}; 
solid curves: compilations of data by \citet{janz-et-al1969} for amorphous $\mathrm{SiO_2}$ (left) and \citet{iapws1994} for $\mathrm{H_2O}$ (right); dashed lines: linear extrapolations of the compiled data. \label{figa2}}
\end{figure}

In spite of its importance, the surface energy of a solid that is composed of a material of astronomical interest is not always available in the literature.
It is not often that we can obtain experimental data on the surface energy of a solid, if the solid matter is a liquid or a gas under Earth's ambient pressure and temperature.
When experimental data on the surface energy are unavailable, we may derive the surface energy $\gamma$ of a solid from the surface tension $\sigma$ of a liquid using the following relationship \citep{ip-toguri1994}
\begin{linenomath*}
\begin{equation}
\gamma = \sigma - T\,\frac{d\,\sigma}{d\,T} , \label{tension}
\end{equation}
\end{linenomath*}
where $T$ is the temperature of the liquid.
The surface tension $\sigma$ usually decreases with the temperature (i.e., $d\,\sigma / d\,T < 0$) and vanishes at the critical temperature $T_\mathrm{c}$.
As long as the surface tension is expressed by a linear equation $\sigma = a + b\,T$ with $a$ and $b$ being fit coefficients, we have $\sigma = \gamma$ at $T=0$ \citep{eoetvoes1886,palit1956}.
To check the validity of Eq.~(\ref{tension}), we plot the temperature dependences of surface tension measured (solid lines) for (a) $\mathrm{SiO_2}$ and (b) $\mathrm{H_2O}$ as well as their linear extrapolations (dashed lines) to lower temperature in Fig.~\ref{figa2} \citep{janz-et-al1969,iapws1994}.
Interestingly, the temperature coefficient of $\mathrm{SiO_2}$ is positive (i.e., $d\,\sigma / d\,T > 0$) in the range of $1773~\mathrm{K} < T < 2080~\mathrm{K}$, but the surface tension is expected to decrease with temperature at higher temperature ($T \ga 2000~\mathrm{K}$) and vanish at the critical temperature $T_\mathrm{c}$ \citep{kraus-et-al2012}\footnote{\citet{kraus-et-al2012} adopted $T_\mathrm{c} = 5130~\mathrm{K}$ for $\mathrm{SiO_2}$, but the critical temperature of $\mathrm{SiO_2}$ is still open to debate in the range of $T_\mathrm{c} = 4862$--$13500~\mathrm{K}$ \citep{iosilevskiy-et-al2013}.}.
The surface energies $\gamma_\mathrm{SiO_2} = 243.08 \pm 0.12~\mathrm{mJ~m^{-2}}$ for $\mathrm{SiO_2}$ and $\gamma_\mathrm{H_2O} = 114.45 \pm 0.07~\mathrm{mJ~m^{-2}}$ for $\mathrm{H_2O}$ ice derived from the extrapolations agree with the values in the literature: $\gamma \approx 250~\mathrm{mJ~m^{-2}}$ for amorphous silica \citep{kimura-et-al2015}; $\gamma \approx 110~\mathrm{mJ~m^{-2}}$ for water ice \citep{israelachvili2011}.

\section{Effect of Adsorption on Surface Energy}
\label{adsorption}

When the surface of a bulk is covered by a thin layer of material, then the surface energy $\gamma$ can be approximately given by the equation that is expressed as \citep{israelachvili1972,israelachvili2011}:
\begin{linenomath*}
\begin{equation}
\gamma = ({\sqrt{\gamma_\mathrm{1}} - \sqrt{\gamma_\mathrm{2}}})^2 , \\
\label{contamination}
\end{equation}
\end{linenomath*}
where $\gamma_\mathrm{1}$ and $\gamma_\mathrm{2}$ are the surface energies of the bulk and the thin layer, respectively.
To check the validity of Eq.~(\ref{contamination}), we consider the case that the surface of amorphous silica is covered by water molecules under atmospheric condition.
If we insert $\gamma_\mathrm{1} = \gamma_\mathrm{SiO_2}$ and $\gamma_\mathrm{2} = \gamma_\mathrm{H_2O}$ into Eq.~(\ref{contamination}), we obtain $\gamma = 0.024~\mathrm{J~m^{-2}}$, which is consistent with $\gamma = 0.025~\mathrm{J~m^{-2}}$ measured for hydrophilic amorphous silica under ambient conditions \citep{kendall-et-al1987}.
Accordingly, we may apply Eq.~(\ref{contamination}) to the determination of the surface energy for carbonaceous matter covered by a thin layer of indium.
On the assumption that the surface energy of indium is $\gamma_\mathrm{In} = 0.631~\mathrm{J~m^{-2}}$, Eq.~(\ref{contamination}) with $\gamma_\mathrm{1} = \gamma_\mathrm{a\mathchar`-C:H}$ and $\gamma_\mathrm{2} = \gamma_\mathrm{In}$ gives $\gamma = 0.37~\mathrm{J~m^{-2}}$ for indium-covered hydrogenated amorphous carbon \citep[cf.][]{alchagirov-et-al2001,alchagirov-et-al2014}.

\end{document}